\documentclass[prd,aps, floatfix,preprintnumbers,showpacs,nofootinbib,notitlepage,superscriptaddress,10pts ]{revtex4-1}
\usepackage[retainorgcmds]{IEEEtrantools}
\usepackage{amsmath}  
\usepackage{amssymb}  
\usepackage{amsfonts}
\usepackage[usenames,dvipsnames]{color}
\usepackage[pdftex]{graphicx}
\usepackage{indentfirst}
\usepackage[pdftex]{hyperref}
\newcommand{\pa}{\partial}

\newcommand{\be}{\begin{equation}}
\newcommand{\ee}{\end{equation}}
\newcommand{\ba}{\begin{eqnarray}}
\newcommand{\ea}{\end{eqnarray}}

\newcommand{\en}{\nonumber\\}

\newcommand{\de}{\delta}

\newcommand{\kk}{\mathbf{k}}
\newcommand{\xx}{\mathbf{x}}

\newcommand{\p}{\mathbf{p}}
\newcommand{\q}{\mathbf{q}}
\newcommand{\pp}[1]{\mathbf{p}_{#1}}
\newcommand{\qq}{\mathbf{q}}

\definecolor{darkred}{RGB}{175,0,0}
\definecolor{darkblue}{RGB}{0,0,175}
\newcommand{\fD}{f_{\chi}}
\newcommand{\fR}{f_{\rm DR}}
\newcommand{\fc}{f_{\chi}}

\newcommand{\mnras}{Mon. Not. R. Astron. Soc.}

\newcommand{\jcap}{J. Cosmol. Astropart. Phys.}
\begin{document}

\title{ETHOS - An Effective Theory of Structure Formation: \\ From dark particle physics to the matter distribution of the Universe}

\author{Francis-Yan Cyr-Racine}\email{fcyrraci@physics.harvard.edu}
\affiliation{Department of Physics, Harvard University, Cambridge, Massachusetts 02138, USA}
\affiliation{California Institute of Technology, Pasadena, California 91125, USA}

\author{Kris Sigurdson}
\affiliation{School of Natural Sciences, Institute for Advanced Study, Princeton, New Jersey 08540, USA}
\affiliation{Department of Physics and Astronomy, University of British Columbia, Vancouver, British Columbia V6T 1Z1, Canada}

\author{Jes\'us Zavala}
\affiliation{Dark Cosmology Centre, Niels Bohr Institute, University of Copenhagen, 2100 Copenhagen, Denmark}

\author{Torsten Bringmann}
\affiliation{Department of Physics, University of Oslo, Box 1048 NO-0316 Oslo, Norway} 

\author{Mark Vogelsberger}
\affiliation{Department of Physics, Kavli Institute for Astrophysics and Space Research, Massachusetts Institute of Technology, Cambridge, Massachusetts 02139, USA}

\author{ Christoph Pfrommer}
\affiliation{Heidelberg Institute for Theoretical Studies, Schloss-Wolfsbrunnenweg 35, D-69118 Heidelberg, Germany}

\date{\today}
%

%%%
\begin{abstract}
We formulate an effective theory of structure formation (ETHOS) that enables cosmological structure formation to be computed in almost any microphysical model of dark matter physics.
This framework maps the detailed microphysical theories of particle dark matter interactions into the physical effective parameters that shape the linear matter power spectrum and the self-interaction transfer cross section of nonrelativistic dark matter.  These are the input to structure formation simulations, which follow the evolution of the cosmological and galactic dark matter distributions. Models with similar effective parameters in ETHOS but with different dark particle physics would nevertheless result in similar dark matter distributions. We present a general method to map an ultraviolet complete or effective field theory of low-energy dark matter physics into parameters that affect the linear matter power spectrum and carry out this mapping for several representative particle models.  We further propose a simple but useful choice for characterizing the dark matter self-interaction transfer cross section that parametrizes self-scattering in structure formation simulations. Taken together, these effective parameters in ETHOS allow the classification of dark matter theories according to their structure formation properties rather than their intrinsic particle properties, paving the way for future simulations to span the space of viable dark matter physics relevant for structure formation.   
\end{abstract}

\maketitle

%%%%%%
\section{Introduction}\label{Introduction}
%%%%%%

Dark matter forms the vast majority of the  matter density in our Universe and plays a crucial role in determining the characteristics of astrophysical structures from galactic to cosmological length scales. Through its gravitational influence on standard baryonic matter, dark matter largely controls the formation and evolution of luminous objects such as galaxies and clusters  \citep{1981ApJ...250..423D,Blumenthal:1982mv,Blumenthal:1984bp,Davis:1985rj}. Within the standard model of structure formation, it is assumed that dark matter is fully nonrelativistic and interacts purely via gravitational interactions in later epochs of the Universe so that gravity is the only dark matter interaction relevant to  the physics of galaxies. However, a significant dark matter thermal velocity dispersion \citep{Bond:1983hb,Colombi:1995ze,SommerLarsen:1999jx,Bode:2000gq,Dalcanton:2000hn,Boyanovsky:2011aa,Boyanovsky:2011ab} or the presence of nongravitational interactions in the dark matter sector such as self-interactions \citep{Goldberg:1986nk,1992ApJ...398...43C,1992ApJ...398..407G,1994ApJ...431...41M,1995ApJ...452..495D,AtrioBarandela:1996ur,Spergel:1999mh,Hannestad:2000gt,AtrioBarandela:2000aw,Stiele:2010xz,Tulin:2012wi,Tulin:2013teo,Kaplinghat:2013xca,Kaplinghat:2015aga}, coupling to Standard Model particles \citep{Green:2005fa,Profumo:2006bv,Bringmann:2009vf,Boehm:2001hm,Mangano:2006mp,Serra:2009uu,McDermott:2010pa,Aarssen:2012fx,Wilkinson:2013kia,Wilkinson:2014ksa,Dvorkin:2013cea,Boehm:2014vja,Escudero:2015yka} or to other yet unknown dark sector particles \citep{Foot:2004pa,Ackerman:2008gi, ArkaniHamed:2008qn,Feng:2009mn,Kaplan:2009de,Behbahani:2010xa,Kaplan:2011yj,Aarssen:2012fx,Cline:2012is,Hooper:2012cw,Das:2012aa,Cyr-Racine:2013ab,Diamanti:2012tg,Baldi:2012ua,Fan:2013yva,Fan:2013tia,McCullough:2013jma,Cyr-Racine:2013fsa,Cline:2013pca,Cline:2013zca,Bringmann:2013vra,Chu:2014lja,Archidiacono:2014nda,Randall:2014kta,Buen-Abad:2015ova,Lesgourgues:2015wza,Choquette:2015mca}, can significantly alter the distribution of dark matter at early times and affect the structure and evolution of astrophysical structures. Given the very large number of possible dark matter theories with these nonstandard characteristics, it is important to identify which of their properties have the largest impact on the structure of the Universe at low redshift. This is a challenging endeavor since the length scales impacted by allowed nonstandard dark matter physics lie deep in the nonlinear regime and their evolution must be computed through expensive numerical simulations \citep{Vogelsberger:2012ku,Zavala:2012us,Vogelsberger:2012sa,Rocha:2012jg,Peter:2012jh,Kamada:2013sh,Vogelsberger:2014pda,Buckley:2014hja,Schewtschenko:2014fca}. To address this situation, we develop here an ``Effective THeory Of Structure formation" (ETHOS), in which the dark matter microphysics is systematically mapped to effective parameters that directly control the formation and evolution of structures. The usefulness of such a framework is clear: all dark matter particle models that map to a given effective theory can be simultaneously constrained by comparing a single ETHOS numerical simulation to observations at no extra cost or effort. We caution the reader that ETHOS is not an ``effective theory'' in the usual sense (i.e. in the context of theoretical particle physics) as it does not involve integrating out ultraviolet physics to get a low-energy description of a theory. It rather represents a grouping of dark matter particle theories into broad categories whose structure formation histories can be described by a handful of ``effective'' ETHOS parameters. 

While the particle physics of potential dark matter candidates is rich and varied, only certain key characteristics of dark matter particles are relevant to structure formation.
The goal of ETHOS is to provide a convenient parametrization of the dark matter physics that matters most to structure formation on a broad range of astrophysical scales. Since we are primarily concerned with the potential presence of new nongravitational interactions in the dark sector and their impact on structure formation, we focus here on models where dark matter can couple to a relativistic component prior to matter-radiation equality \emph{and} have significant self-interaction inside halos today (see, e.g., \cite{Aarssen:2012fx,Cyr-Racine:2013ab}). Within this type of theories, the ETHOS parametrization that we develop in the present work provides a nearly universal language to translate dark matter particle physics models into quantities that directly affect how structures assemble and evolve in our Universe. The ETHOS language is particularly useful when comparing dark matter models with observational data. For cosmologists and astrophysicists, ETHOS yields an accessible and easy-to-use framework to study deviations from the pure cold dark matter scenario, without the need to delve into the details of dark matter particle models. For particle physicists, ETHOS provides a clear relation between the dark matter microphysics and its impact on structure formation at multiple scales. As part of the ETHOS project, we make publicly available a Boltzmann code\footnote{The code is publicly available at \url{https://bitbucket.org/franyancr/ethos_camb}.} (based on the cosmological code \texttt{CAMB} \citep{Lewis:2002ah}) that allows the computation of the linear dark matter transfer function for a broad range of dark matter models.     
  
In this  paper we introduce the ETHOS framework and present the structure-formation-focused parametrization we use in the linear regime to describe dark matter models that have significant nongravitational interactions. A large portion of this paper is devoted to characterizing how the microphysics describing the dark sector is ultimately responsible for the exact structure of the dark matter transfer function. We also discuss a simple approach to capture aspects of the self-interaction transfer cross section's velocity dependence which are most relevant for structure formation. In a companion paper \citep{2016MNRAS.460.1399V}, we present a suite of high-resolution zoom simulations that explore the structure of a Milky-Way-type galaxy in a few ETHOS scenarios motivated by the model described in Refs.~\cite{Aarssen:2012fx,Bringmann:2013vra}. These simulations explore, for the first time 
in a consistent model of particle physics and at this resolution, the joint impact of a nonstandard initial spectrum of dark matter fluctuations and significant self-interactions inside halos today.

While we focus here on dark matter models that have significant nongravitational interactions, we emphasize that the ETHOS concept is much broader than this particular family of scenarios and could eventually be expanded to include other types of dark matter physics, such as warm or decaying dark matter (see e.g. \cite{Huo:2011nz,Aoyama:2014tga,Audren:2014bca,Wang:2014ina}). We also note that the parametrization introduced here is general enough to approximately capture dark matter physics that would naively appear impossible to be describable within our current implementation of ETHOS. For instance, as was discussed in Ref.~\cite{Buckley:2014hja} (see also Ref.~\cite{2016MNRAS.460.1399V}), dark matter models displaying strongly damped acoustic oscillations in their initial matter power spectrum lead to a structure formation similar to that of warm dark matter (assuming a negligible level of self-interaction).  We therefore expect the framework presented here to be very useful to describe several types of departures from the standard cold dark matter theory.

The structure of this paper is as follows.  Section \ref{sec:linear_mat_power} is entirely devoted to presenting and studying the physics and equations determining the structure of the linear matter power spectrum. In particular, Sec.~\ref{sec:Gen_Boltz_eqs_sum} summarizes the key equations and quantities necessary to compute the matter power spectrum. The full derivation of these equations is presented in Appendix \ref{sec:Gen_Boltz_eqs}. In Sec.~\ref{sec:matter_power_recipe}, we introduce the ETHOS parametrization and show several examples of how the mapping from particle physics parameters to effective ETHOS parameters is done in practice. In Sec.~\ref{sec:power_shape}, we illustrate how the dark matter transfer function depends on the most relevant ETHOS parameters. In Sec.~\ref{sec:self-interaction}, we present our parametrization for the velocity dependence of the dark matter self-interaction transfer cross section and illustrate its usefulness with some examples. We summarize the main points and motivations of the ETHOS framework in Sec.~\ref{sec:ETHOS}, and finally conclude in Sec.~\ref{sec:conclusions}. 

Unless otherwise noted, we assume throughout a spatially flat universe with 3.046 massless neutrinos with the following standard cosmological parameters: baryon density $\Omega_{\rm b}h^2 = 0.02197$, dark matter density $\Omega_{\rm DM} h^2 = 0.12206$, Hubble constant $H_0 = 69.09$ km/s/Mpc, power spectrum amplitude $A_{\rm s} = 2.1758\times10^{-9}$, spectral index $n_{\rm s} = 0.9671$, and optical depth $ \tau_{\rm reio} = 0.089$. 
%%%%
\section{The linear power spectrum for dark matter}\label{sec:linear_mat_power}
%%%%
One of the most important quantities determining the structure formation history within a given dark matter scenario is the matter power spectrum which characterizes the amplitude of matter fluctuations on different scales. At early times, the matter power spectrum can be computed by solving the linearized equations for the evolution of the matter (both dark and baryonic) density and velocity.\footnote{In the case of relativistic dark matter, the shear and higher moments of the dark matter Boltzmann equation must also be evolved.} As matter perturbations grow, they eventually enter the nonlinear regime and other methods (i.e. numerical simulations, see Ref.~\cite{2016MNRAS.460.1399V}) must be used to compute the power spectrum. The \emph{linear} matter power spectrum is nonetheless a very useful quantity since it provides approximate guidelines about the smallest possible bound structures that can form within any dark matter scenario and is used to set the initial conditions for numerical simulations. 

In this first ETHOS paper, we focus on a scenario in which a single species of dark matter (DM, denoted by $\chi$) can interact with a relativistic component (denoted by $\tilde{\gamma}$) which we will generally refer to as ``dark radiation'' (DR) but could also be made of Standard Model neutrinos or photons. We consider the situation where the only relevant process\footnote{We note that elastic DM self-interaction $\chi\chi\leftrightarrow\chi\chi$ is irrelevant for the cosmological evolution of {\it linear} perturbations, unless the DM is itself relativistic. See Appendix \ref{sec_app:chichi_int} for details.} for DM is its 2-to-2 scattering with DR, $\chi\tilde{\gamma}\leftrightarrow\chi\tilde{\gamma}$, but allow for DR self-interactions through the process $\tilde{\gamma}\tilde{\gamma}\leftrightarrow\tilde{\gamma}\tilde{\gamma}$. We assume that the DM relic abundance is fixed at some high temperature (through e.g. thermal freeze-out) and we therefore neglect here the effect of DM annihilation or decay on the evolution of DM fluctuations. We note however that these latter processes could be included in future versions of the ETHOS framework. 

In this section, our goal is to describe how the nonstandard DM physics enters the computation of the linear matter power spectrum. Since we are mainly interested in the impact of this nontrivial DM physics on structure formation, we focus our attention exclusively on scalar cosmological fluctuations and leave the study of tensor fluctuations to future work. We present in Appendix \ref{sec:Gen_Boltz_eqs} a detailed derivation of the coupled equations describing the evolution of DM and DR perturbations. In the following, we shall first summarize the key results from that Appendix before describing a general procedure to compute the linear matter power spectrum within the ETHOS framework.
\subsection{Dark Matter and Dark Radiation Perturbation Equations}\label{sec:Gen_Boltz_eqs_sum}
In the following section, we summarize the key results from Appendix \ref{sec:Gen_Boltz_eqs}. We invite the interested reader to consult that Appendix for more details. Our goal here is to obtain the equations of motion for the DM and DR density perturbations, denoted by $\de_\chi$ and $\de_{\rm DR}$, respectively. These equations must be solved together with those describing the evolution of baryons, photons, and neutrinos in order to compute the linear matter power spectrum (see e.g.~Ref.~\cite{Ma:1995ey}). In the following, we assume that DM is made of massive, highly nonrelativistic particles interacting with a massless DR component. For these choices, the momentum transferred in a typical DM-DR collision is small, which dramatically simplifies the computation of the collision integral (see Sec.~\ref{app:sec:First-order collision term} of Appendix \ref{sec:Gen_Boltz_eqs}). We further assume the DR to have a thermal spectrum. In conformal Newtonian gauge, the equations describing the evolution of DR perturbations are
\ba
\dot{\de}_{\rm DR} +\frac{4}{3}\theta_{\rm DR}-4\dot{\phi}&=&0,\label{eq:DR_continuity_sum}\\
\dot{\theta}_{\rm DR}+k^2(\sigma_{\rm DR}-\frac{1}{4}\de_{\rm DR})-k^2\psi&=&\dot{\kappa}_{\rm DR-DM} \,(\theta_{\rm DR}-\theta_{\chi}),\label{eq:DR_Euler_sum}\\
\dot{\Pi}_{{\rm DR},l} + \frac{k}{2l+1}\left((l+1)\Pi_{{\rm DR},l+1} - l\Pi_{{\rm DR},l-1}\right)&=&\left(\alpha_l\dot{\kappa}_{\rm DR-DM} + \beta_l \dot{\kappa}_{\rm DR-DR}\right)\,\Pi_{{\rm DR},l},\label{eq:DR_higher_l_moments_sum}
\ea
where $\theta_\chi\equiv i \kk \cdot \vec{v}_{\chi}$ is the divergence of the DM bulk velocity in Fourier space,  $\theta_{\rm DR} $ is the divergence of the DR velocity in Fourier space, $\phi$ and $\psi$ are the two gravitational potentials in the conformal Newtonian gauge, $\sigma_{\rm DR}$ is the DR shear stress, $k=|\kk|$ is the comoving wave number of the perturbation, $\Pi_{{\rm DR},l}$ is the $l^{\rm th}$ moment of the DR multipole hierarchy, $\dot{\kappa}_{\rm DR-DM}$ is the DR opacity to DM scattering, which is given by
\be\label{eq:dark_rad_opacity_def_sum}
\dot{\kappa}_{\rm DR-DM} =\left(\frac{3n_\chi^{(0)}m_\chi}{4\rho_{\rm DR}}\right)\frac{a}{16\pi m_\chi^3}   \frac{\eta_{\rm DR}}{3}\int \frac{p^2dp}{2\pi^2}\, p^2 \frac{\pa\fR^{(0)}(p)}{\pa p}\left[ A_0(p) - A_1(p)\right]  ,
\ee
where the homogeneous part of the DR energy density is $\rho_{\rm DR} = \eta_{\rm DR} \zeta \pi^2 T_{\rm DR}^4/30$ with $\zeta=1$ for bosonic DR and $\zeta=7/8$ for fermionic DR, $a$ is the cosmological scale factor, $p$ is the magnitude of the three-momentum, $m_\chi$ is the DM mass, $n_\chi^{(0)}$ is the spatially homogeneous DM number density, $T_{\rm DR}$ is the temperature of the DR, $\fR^{(0)}$ is the homogeneous part of the DR phase-space density, and where the $A_l$ coefficients are the projection of the spin-summed squared matrix element onto the $l^{\rm th}$ Legendre polynomial $P_l(x)$
\be\label{eq:A_l_coefficients_sum}
A_l(p) = \frac{1}{2}\int_{-1}^1 d\tilde{\mu}\, P_l(\tilde{\mu}) \left(\frac{1}{\eta_\chi\eta_{\rm DR}}\sum_{\rm states} |\mathcal{M}|^2\right)\Bigg{|}_{\begin{subarray}{l} t=2p^2(\tilde{\mu}-1) \\ s=m_\chi^2+2p m_\chi \end{subarray}}.
\ee
In the above, $\eta_\chi$ and $\eta_{\rm DR}$ are the DM and DR spin and color degeneracy factors, respectively, and $|\mathcal{M}|^2$ is the square of the matrix element for the $\chi\tilde{\gamma}\leftrightarrow\chi\tilde{\gamma}$ process written in terms of the Mandelstam variables $s$ and $t$. Throughout, an overhead dot denotes a derivative with respect to conformal time. In Eq.~\eqref{eq:DR_higher_l_moments_sum}, the coefficients $\alpha_l$ are $l$-dependent factors that encompass information about the angular dependence of the DM-DR scattering cross section. They are given by
\be\label{eq:def_ang_coefficients_DR_sum}
\alpha_l \equiv \frac{\int dp\, p^4 \frac{\pa\fR^{(0)}(p)}{\pa p}\left[A_0(p) -A_l(p)\right]}{\int dp\, p^4 \frac{\pa\fR^{(0)}(p)}{\pa p}\left[A_0(p) -A_1(p)\right]}.\ee
In models where DR self-interaction is allowed, the function $\dot{\kappa}_{\rm DR-DR}$ appearing in Eq.~\eqref{eq:DR_higher_l_moments_sum} is the opacity for that process and $\beta_l$ are the corresponding angular coefficients [see Eqs.~\eqref{eq:dark_rad_self_opacity_def} and \eqref{eq:def_ang_coefficients_DR-DR} for more details]. 

The equations governing the DM perturbations are
\ba
\dot{\de}_\chi+\theta_\chi-3\dot{\phi}&=&0\label{eq:DM_continuity_right_not_sum},\\
\dot{\theta}_\chi -c_{\chi}^2k^2\de_\chi+\mathcal{H} \theta_\chi - k^2\psi &=&\dot{\kappa}_\chi \left[\theta_\chi -\theta_{\rm DR} \right]\label{eq:DM_euler_approx_sec_sum},
\ea
where $c_\chi$ is the adiabatic DM sound speed, $\mathcal{H}$ is the conformal Hubble rate, and $\dot{\kappa}_\chi$ is the DM drag opacity. The latter is given by
\be\label{eq:def_DM_drag_opacity_sum}
 \dot{\kappa}_\chi  = a\frac{1}{16\pi m_\chi^3} \frac{\eta_{\rm DR} }{3} \int \frac{p_1^2dp_1}{2\pi^2} p_1^2 \frac{\pa \fR^{(0)}(p_1)}{\pa p_1} \left(A_0(p_1)-A_1(p_1)\right) =\frac{4 \rho_{\rm DR}}{3 n_\chi^{(0)} m_\chi} \dot{\kappa}_{\rm DR-DM}.
\ee
The adiabatic DM sound speed appearing in Eq.~\eqref{eq:DM_euler_approx_sec_sum} is approximately given by
\be\label{eq:DM_sound_speed_final_sum}
c_{\chi}^2 =  \frac{T_\chi}{m_\chi}\left(1-\frac{\dot{T}_\chi}{3 \mathcal{H}  T_\chi}\right),
\ee
where $T_\chi$ is the DM temperature. The evolution of the latter is controlled by 
\be\label{eq:DM_temperature_evol_final_sum}
\frac{dT_\chi}{d\tau} =-2 \mathcal{H} T_\chi + \Gamma_{\rm heat}(T_{\rm DR})\left(T_{\rm DR} - T_\chi\right)\,.
\ee
Here, $\Gamma_{\rm heat}$ stands for the DM heating rate, which can be written as \cite{2007JCAP...04..016B} 
\be\label{eq:Gamma_heating_sum}
\Gamma_{\rm heat}(T_{\rm DR}) = a \frac{\eta_{\rm DR}m_\chi}{6 (2\pi)^3}  \left[\sum_n c_n (n+4)!\zeta(n+4)\gamma_n \left(\frac{T_{\rm DR}}{m_\chi}\right)^{n+4}\right],
\ee
where the coefficients $c_n$ are defined from the matrix element for the $\chi\tilde{\gamma}\leftrightarrow\chi\tilde{\gamma}$ process
\be\label{eq:momentum_power_law_sum}
\left(\frac{1}{\eta_\chi\eta_{\rm DR}}\sum_{\rm states} |\mathcal{M}|^2\right)\Bigg{|}_{\begin{subarray}{l} t=0 \\ s=m_\chi^2+2q m_\chi \end{subarray}} \equiv \sum_n c_n\left(\frac{q}{m_\chi}\right)^n,
\ee
where $q = a p$ is the comoving momentum of the incoming DR, and where $\gamma_n = (1-2^{-n-3})$ for fermionic DR and $\gamma_n=1$ for bosonic DR. In Eq.~\eqref{eq:Gamma_heating_sum}, $\zeta(z)$ is the Riemann Zeta function. We observe that the particle physics details of an interacting DM and DR model only enter through the opacity functions $\dot{\kappa}_\chi$, $\dot{\kappa}_{\rm DR-DM}$ and $\dot{\kappa}_{\rm DR-DR}$, and through the coefficients $\alpha_l$ and $\beta_l$ which depends on the angular dependence of the DM-DR and DR-DR scattering amplitude, respectively. There is also a small dependence on the DM sound speed $c_\chi$, but since it is very small for highly nonrelativistic DM, it plays only a minor role in determining the evolution of the DM density fluctuations unless the wave number $k$ is very large. We now have all the key ingredients necessary to compute the linear matter power spectrum.
\subsection{A general procedure for computing the linear matter power spectrum}\label{sec:matter_power_recipe}
In the previous section (see also Appendix  \ref{sec:Gen_Boltz_eqs}), we have presented the cosmological perturbation equations for a model in which nonrelativistic DM couples to a relativistic component via the process $\chi\tilde{\gamma}\rightarrow\chi\tilde{\gamma}$. While the calculation can become tedious, it suggests a simple recipe to derive the required system of equations: 

\begin{enumerate}
\item For the process   $\chi\tilde{\gamma}\rightarrow\chi\tilde{\gamma}$, compute the spin-summed matrix element squared and evaluate it at $t= 2p^2(1-\tilde{\mu})$ and $s = m_\chi^2+2 p m_\chi$, where $p$ is the momentum of the incoming DR and $\tilde{\mu}$ is the cosine of the angle between the incoming and outgoing DR particle.
\item Compute the $A_l$ coefficients using the projection integral given in Eq.~\eqref{eq:A_l_coefficients_sum}.
\item Compute $\dot{\kappa}_{\rm DR-DM}$ and $\dot{\kappa}_\chi$ using Eqs.~\eqref{eq:dark_rad_opacity_def_sum} and \eqref{eq:def_DM_drag_opacity_sum}, respectively. Compute the angular coefficients $\alpha_l$ using Eq.~\eqref{eq:def_ang_coefficients_DR_sum}.
\item If relevant for the model at hand, compute the opacity $\dot{\kappa}_{\rm DR-DR}$ and the $\beta_l$ coefficients using Eqs.~\eqref{eq:dark_rad_self_opacity_def} and \eqref{eq:def_ang_coefficients_DR-DR}, respectively.
\item Solve Eq.~\eqref{eq:DM_temperature_evol_final_sum} to obtain the DM temperature evolution. Compute the DM adiabatic sound speed $c_\chi^2$ using Eq.~\eqref{eq:DM_sound_speed_final_sum}.
\item Solve Eqs.~\eqref{eq:DR_continuity_sum}-\eqref{eq:DR_higher_l_moments_sum}, \eqref{eq:DM_continuity_right_not_sum}, and \eqref{eq:DM_euler_approx_sec_sum} using a standard Boltzmann solver in order to obtain the matter power spectrum.
\end{enumerate}
This procedure is straightforward but is not fully amenable to a simple numerical implementation since one would need to code the specific functions $\dot{\kappa}_{\rm DR-DM}$, $\dot{\kappa}_{\rm DR-DR}$, $\kappa_\chi$, and $\Gamma_{\rm heat}$ for each model. While this is in principle possible, one can further simplify the computation by noting that the opacities and heating rate are often power-law functions of the temperature (or redshift). This behavior occurs because the matrix elements entering the collision integrals are often themselves power laws of momentum (see e.g. Eq.~\eqref{eq:momentum_power_law_sum}). We can then write
\be\label{eq:opacities_expansion}
\dot{\kappa}_{\rm DR-DM} = -(\Omega_\chi h^2)x_\chi(z) \sum_n a_n \left(\frac{1+z}{1+z_{\rm D}}\right)^n,\qquad \dot{\kappa}_\chi =-\frac{4}{3} (\Omega_{\rm DR}h^2) x_\chi(z) \sum_n a_n \frac{(1+z)^{n+1}}{(1+z_{\rm D})^n},
\ee
\be\label{eq:opacities_expansion2}
\dot{\kappa}_{\rm DR-DR} = - (\Omega_{\rm DR}h^2) x_{\rm DR-DR}(z) \sum_n b_n \left(\frac{1+z}{1+z_{\rm D}}\right)^n, \qquad \Gamma_{\rm heat}  = (\Omega_{\rm DR}h^2)  x_\chi(z) \sum_n d_n \frac{(1+z)^{n+1}}{(1+z_{\rm D})^n},
\ee
where $a_n$, $b_n$, and $d_n$ are constants with units of inverse length, $h$ is the dimensionless Hubble constant $h = H_0/(100\,{\rm km/s/Mpc})$, $\Omega_\chi$ and $\Omega_{\rm DR}$ are  respectively the DM and DR densities in units of the critical density of the Universe, and where we have introduced the dimensionless functions $x_\chi(z)$ and $x_{\rm DR-DR}(z)$ to take into account possible departures from a pure power-law behavior in some models\footnote{A good example of deviation from pure power-law scaling occurs in the atomic dark matter model at the epoch of dark recombination \citep{Cyr-Racine:2013ab}. Even in this case however, the opacities can generally still be approximated by a (steep) power law close the DM drag epoch.}. In many instances, the physics responsible for nontrivial values of $x_\chi$ and $x_{\rm DR-DR}$ can be computed independently of the $\chi\tilde{\gamma}\rightarrow\chi\tilde{\gamma}$ scattering process considered here, and the above factorization is therefore physically motivated. We have also introduced the redshift $z_{\rm D}$ which is used to normalize the values of the coefficients $a_n$, $b_n$, and $d_n$. The value of $z_{\rm D}$ is arbitrary but choosing it to be the redshift when the DM opacity becomes equal to the conformal Hubble rate $\mathcal{H}$ prevents artificially large or small values for the coefficients defining the opacity and heating expansions. In this work, we choose $z_{\rm D} = 10^7$, which corresponds to a decoupling temperature close to $T_{\rm DR} \sim 1$ keV (assuming $\xi = 0.5$). 

We note that we have written the DM opacity $\dot{\kappa}_\chi$ as an expansion in a term that goes as $(1+z)^{n+1}$ since we typically have $\dot{\kappa}_\chi \propto (1+z)\dot{\kappa}_{\rm DR-DM}$. The factor $4/3$ appearing in this expansion enforces momentum conservation in DM-DR scattering. We also note that the coefficients $a_n$, $b_n$, and $d_n$ are independent of the standard $\Lambda$CDM parameters and thus only depend on the physics of the dark sector. In many models of interest, only a single term in the expansions given in Eqs.~\eqref{eq:opacities_expansion} and \eqref{eq:opacities_expansion2} is nonvanishing. Furthermore, even in more complex cases with multiple nonzero terms or nontrivial $x_\chi(z)$, we expect the opacity and heating rates to be well approximated by a single, though not necessarily integer, power law. 

With these expansions, we now have a clear and straightforward mapping between the couplings, masses, and temperatures defining a given DM particle physics model, and the effective parameters controlling the shape of the linear matter power spectrum. It is important to realize that our parametrization in terms of $a_n$ and $d_n$ coefficients has a clear physical interpretation. Indeed, the presence of nonzero $a_n$ and $d_n$ coefficients directly corresponds to a DM-DR scattering process with a squared matrix element whose behavior is given by
\be
|\mathcal{M}|^2_{\chi\tilde{\gamma}\rightarrow\chi\tilde{\gamma}}\propto \left(\frac{p_{\rm DR}}{m_\chi}\right)^{n-2},
\ee
where $p_{\rm DR}$ is the incoming DR momentum. Assuming that the opacities are pure power laws of redshift [implying $x_{\rm DR-DR}(z)=x_\chi(z)=1$], we schematically have
\be
\Big\{m_\chi, \{g_i\},\{h_i \},\xi \Big\} \rightarrow \Big\{\omega_{\rm DR}, \{a_n,\alpha_l\},\{b_n,\beta_l\}, \{d_n, m_\chi,\xi\}\Big\}\rightarrow P_{\rm lin.,~matter}(k),
\ee
where $\omega_{\rm DR} \equiv \Omega_{\rm DR}h^2$, $\{g_i\}$ represents the set of coupling constants appearing in a given dark matter model, $\{h_i\}$ is a set of other internal parameters such as mediator mass and number of internal degrees of freedom, and we remind the reader that $\xi = (T_{\rm DR}/T_{\rm CMB})|_{z=0}$. With this latter definition, the physical DR energy density today is given by $\omega_{\rm DR} = (\eta_{\rm DR}/2)\zeta\xi^4\Omega_{\gamma}h^2 \simeq 1.235\times10^{-5} \zeta \eta_{\rm DR} \xi^4$, where $\Omega_\gamma$ is the energy density in photons today in unit of the critical density of the Universe, and where $\zeta=1$ for bosonic DR and $\zeta=7/8$ for fermionic DR. Current temperature and polarization measurements of the cosmic microwave background by the Planck satellite \cite{2015arXiv150201589P} constrain the energy density in DR to be $\omega_{\rm DR} < 2\times10^{-6}$ at $95\%$ confidence level. 

 From a practical perspective, the above effective parametrization allows us to simplify the computation of the matter power spectrum by directly passing the constant coefficients $\Big\{\omega_{\rm DR}, \{a_n,\alpha_l\},\{b_n,\beta_l\}, \{d_n, m_\chi,\xi\}\Big\}$ to a Boltzmann code, without having to hard code the functional form of the DM and DR opacities for each particle model. For this purpose, we have modified the Boltzmann code \texttt{CAMB} \citep{Lewis:2002ah} in order to pass to it the array of effective ETHOS parameters. This code is publicly available at \url{https://bitbucket.org/franyancr/ethos_camb}. 

We emphasize that not all effective parameters have a large impact on the matter power spectrum. For instance, the subset $\{d_n, m_\chi,\xi\}$ is only used to determine the small DM adiabatic sound speed. Thus, these parameters have very little impact on the actual structure of the linear matter power spectrum, except on very small scales. Similarly, the subset $\{b_n,\beta_l\}$ only directly affects the evolution of the DR and will have a subleading effect on the DM distribution.  We do note that parameters like the $a_n$, $b_n$ or $d_n$ can themselves implicitly depend on other physical parameters, such as $\xi$, but we use these coefficients to characterize such dependence.  We leave to future work the detailed study of the impact of subdominant parameters on the matter power spectrum and focus here on the most relevant parameters $ \Big\{\omega_{\rm DR},\{a_n\,,\alpha_l\}\Big\}$. We now illustrate this ETHOS mapping with some concrete examples. 

\subsubsection{DM-DR scattering via a massive mediator}\label{ref_model}
We first consider a model where DM can interact with a massless sterile neutrino ($\nu_{\rm s}$) via a broken $U(1)$ interaction mediated by a massive vector boson $\phi_\mu$ \cite{Aarssen:2012fx,Bringmann:2013vra, Dasgupta:2013zpn}. The interaction Lagrangian is given by
\be
\mathcal{L}_{\rm int} = -g_\chi \phi_\mu\bar{\chi}\gamma^\mu\chi-\frac{1}{2}g_\nu \phi_\mu\bar{\nu}_{\rm s}\gamma^\mu \nu_{\rm s}-\frac{1}{2}m_\phi^2 \phi_\mu\phi^\mu-\frac{1}{2}m_\chi\bar{\chi}\chi,
\ee
in addition to the standard kinetic terms. Here, we have $\eta_\chi = \eta_{\nu_{\rm s}} = 2$. The spin-summed matrix element for the scattering  $\nu_{\rm s}(\pp{1})+\chi(\pp{2})\leftrightarrow \nu_{\rm s}(\pp{3})+\chi(\pp{4}) $ is
\be
\frac{1}{\eta_\chi \eta_{\nu_{\rm s}}  }\sum_{\rm spins}|\mathcal{M}|^2= \frac{2g_\chi^2g_\nu^2}{(m_\phi^2-t)^2}\left( t^2 + 2 s t +2(m_\chi^2-s)^2\right).
\ee
We then evaluate the matrix element in the limit $t= 2p_1^2(\tilde{\mu}-1)$ and $s= m_\chi^2 +2 m_\chi p_1$,
\ba\label{eq:simple_mat_element}
\left(\frac{1}{\eta_\chi  \eta_{\nu_{\rm s}} }\sum_{\rm spins}|\mathcal{M}|^2\right)\Bigg{|}_{\begin{subarray}{l} t=2p_1^2(\tilde{\mu}-1) \\ s=m_\chi^2+2p_1m_\chi \end{subarray}} &=& \frac{8 g_\chi^2g_\nu^2}{\left(m_\phi^2 - 2p_1^2(\tilde{\mu}-1)\right)^2}p_1^2\left( m_\chi^2(1+\tilde{\mu}) +2 m_\chi p_1(\tilde{\mu}-1)+p_1^2(\tilde{\mu}-1)^2\right)\en 
&\rightarrow& \frac{8 g_\chi^2g_\nu^2}{m_\phi^4}m_\chi^2 p_1^2 (1+\tilde{\mu}) \qquad \text{for} \qquad p_1\ll m_\phi < m_\chi
\ea
where we have simplified the result for the case of nonrelativistic DM in the last line. We then obtain
\be
A_0(p_1) = \frac{8 g_\chi^2g_\nu^2}{m_\phi^4}m_\chi^2 p_1^2,\qquad A_1(p_1) = \frac{A_0(p_1)}{3},\qquad A_{l\geq2}(p_1) =0,
\ee
which immediately leads to 
\be
\dot{\kappa}_{\rm DR-DM} = -a \frac{3}{2}\frac{\pi g_\chi^2 g_\nu^2}{m_\phi^4} \left(\frac{310}{441}\right)  n_\chi^{(0)} T_{\rm DR}^2,\qquad \dot{\kappa}_\chi =  \frac{4\rho_{\rm DR}}{3\rho_\chi}\dot{\kappa}_{\rm DR-DM},\qquad \alpha_{l\geq2} = \frac{3}{2}.
\ee
Since $T_{\rm DR}\propto(1+z)$, we finally obtain
\be
a_{n\leq3}=0,\qquad a_4 = (1+z_{\rm D})^{4}\frac{3}{2}\frac{\pi g_\chi^2g_\nu^2}{m_\phi^4} \frac{ \tilde{\rho}_{\rm crit} }{ m_\chi} \left(\frac{310}{441}\right) \xi^2 T_{\rm CMB,0}^2\,,\qquad a_{n\geq5}=0,
\ee
where $\tilde{\rho}_{\rm crit}  \equiv  \rho_{\rm crit}/h^2 \simeq8.098\times10^{-11}$eV$^4$ is a constant independent of cosmological parameters. The current temperature of the CMB is denoted by $T_{\rm CMB,0}$. We also have that $x_\chi(z) = 1$. Thus, for this model the ETHOS mapping takes the form
\be
\Big\{m_\chi, m_\phi,g_\chi,g_\nu,\xi,\eta_\chi, \eta_{\nu_{\rm s}}  \Big\}\rightarrow \Big\{\omega_{\rm DR}, a_4,\alpha_{l\geq2}=\frac{3}{2}\Big\}.
\ee
\subsubsection{Hidden-charged scalar DM}
We now consider a complex scalar DM candidate charged under a new unbroken dark $U(1)$ interaction mediated by the gauge field $\tilde{A}_\mu$ \citep{Feng:2009mn}. The interaction Lagrangian is given by
\be
\mathcal{L}_{\rm int} = - (D^\mu \chi )^\dagger D_\mu \chi -m_\chi^2 \chi^\dagger\chi, \qquad\text{where}\qquad D_\mu = \pa_\mu - i g_\chi \tilde{A}_\mu.
\ee
Here, we have $\eta_\chi = \eta_{\rm DR} = 2$. The spin-summed matrix element for the scattering  $\tilde{\gamma}(\pp{1})+\chi(\pp{2})\leftrightarrow \tilde{\gamma}(\pp{3})+\chi(\pp{4}) $ is \citep{Feng:2009mn}
\be
\frac{1}{\eta_\chi\eta_{\rm DR}}\sum_{\rm spins}|\mathcal{M}|^2= \frac{4 g_\chi^4\left[(m_\chi^2-s)^4+2(m_\chi^2-s)^2st + (m_\chi^4+s^2)t^2\right]}{(m_\chi^2-s)^2(s+t-m_\chi^2)^2},
\ee
which immediately leads to
\be
\left(\frac{1}{\eta_\chi\eta_{\rm DR}}\sum_{\rm spins}|\mathcal{M}|^2\right)\Bigg{|}_{\begin{subarray}{l} t=2p_1^2(\tilde{\mu}-1) \\ s=m_\chi^2+2p_1m_\chi \end{subarray}} = \frac{2 g_{\chi }^4 \left(\left(\tilde{\mu}^2+1\right) m_\chi^2+2 \left(\tilde{\mu}^2-1\right) m_\chi p_1+2 (\tilde{\mu}
   -1)^2 p_1^2\right)}{(m_\chi+(\tilde{\mu} -1) p_1)^2} \longrightarrow 2 g_\chi^4 (1+\tilde{\mu}^2),
\ee
where we have taken the limit $p_1\ll m_\chi$. The coefficients of the Legendre expansion for the matrix element are then
\be
A_0(p_1) = \frac{8 g_\chi^4}{3},\qquad A_1(p_1) = 0,\qquad A_2(p_1) = \frac{4 g_\chi^4}{15},\qquad A_{l\geq3}(p_1)=0.
\ee
Using Eqs.~\eqref{eq:dark_rad_opacity_def_sum} and \eqref{eq:def_DM_drag_opacity_sum}, the DR and DM drag opacities and the angular coefficients are
\be
\dot{\kappa}_{\rm DR-DM} = -a \frac{ g_\chi^4}{6 \pi m_\chi^2}n_\chi^{(0)} ,\qquad \dot{\kappa}_\chi = \frac{4\rho_{\rm DR}}{3\rho_\chi}\dot{\kappa}_{\rm DR-DM},\qquad \alpha_2 = \frac{9}{10},\qquad \alpha_{l\geq3}=1.
\ee
The astute reader will recognize the above expressions as similar ones arise in the case of CMB photons scattering off free electrons if polarization is neglected. The opacity coefficients are then
\be
a_0 = 0, \quad a_1 =0,\quad a_2 = (1+z_{\rm D})^2 \frac{ g_\chi^4}{6\pi m_\chi^2}\frac{ \tilde{\rho}_{\rm crit}}{m_\chi},\quad a_{n\geq3}=0,
\ee
Here, the ETHOS mapping takes the form
\be
\Big{\{} m_\chi, g_\chi, \xi, \eta_\chi, \eta_{\rm DR} \Big{\}}\rightarrow \left\{\omega_{\rm DR},a_2,\alpha_{2} = \frac{9}{10},\alpha_{l\geq3} =1\right\}.
\ee
\subsubsection{DM coupled to non-Abelian DR}
Here, we focus on the scenario discussed in \citep{Buen-Abad:2015ova, Lesgourgues:2015wza} where DM is a Dirac fermion in the fundamental representation of a dark SU$(N)_{\rm d}$ gauge group. The non-Abelian gauge coupling $g_{\rm d}$ is always assumed to be small such that confinement does not occur. Since DR is always self-interacting in this model, DR forms a perfect fluid and multipoles with $l\geq2$ are strongly suppressed (this is equivalent to setting $\beta_{l\geq2}\dot{\kappa}_{\rm DR-DR} \rightarrow \infty$). The interaction Lagrangian is
\be
\mathcal{L}_{\rm int} = - (D^\mu \chi )^\dagger D_\mu \chi -m_\chi^2 \chi^\dagger\chi, \qquad\text{where}\qquad D_\mu = \pa_\mu - i g_{\rm d} T^a \tilde{A}_\mu^a,
\ee
where $T^a$ are the SU$(N)_{\rm d}$ generators, and $\tilde{A}_\mu^a$ are the non-Abelian gauge fields. Here, we have $\eta_\chi = 2N$ and $\eta_{\rm DR} = 2(N^2-1)$. The matrix element for the $t$-channel scattering  $\tilde{\gamma}_{\rm d}(\pp{1})+\chi(\pp{2})\leftrightarrow \tilde{\gamma}_{\rm d}(\pp{3})+\chi(\pp{4}) $  is \citep{Buen-Abad:2015ova}
\be
\frac{1}{\eta_\chi \eta_{\rm DR}}\sum_{\begin{subarray}{c} \text{spins} \\ \text{colors} \end{subarray}}|\mathcal{M}|^2=\frac{2 g_{\rm d}^4 (s-m_\chi^2)(s+t-m_\chi^2)}{t^2}
\ee
which leads to 
\be
\left(\frac{1}{\eta_\chi \eta_{\rm DR}}\sum_{\begin{subarray}{c} \text{spins} \\ \text{colors} \end{subarray}}|\mathcal{M}|^2\right)\Bigg{|}_{\begin{subarray}{l} t=2p_1^2(\tilde{\mu}-1) \\ s=m_\chi^2+2p_1m_\chi \end{subarray}} = \frac{2 g_{\rm d}^4 m_\chi\left( m_\chi +p_1(\tilde{\mu}-1)\right)}{p_1^2(\tilde{\mu}-1)^2}\longrightarrow \frac{2 g_{\rm d}^4 m_\chi^2}{p_1^2(\tilde{\mu}-1)^2}\qquad\text{for}\qquad p_1\ll m_\chi.
\ee
Since the mediator here is massless, we must regularize the expression for the coefficients $A_l(p_1)$ to avoid the divergence associated with vanishing momentum transfer ($\tilde{\mu}\rightarrow 1$). As in the familiar case of Coulomb scattering, there is a minimum scattering angle $\theta_{\rm min}$ that is directly related to the Debye screening length of the non-Abelian plasma. In terms of this minimum angle, the matrix element coefficients are
\be
A_0(p_1) = \frac{g_{\rm d}^4 m_\chi^2 \cot^2{\left(\frac{\theta_{\rm min}}{2}\right)}}{2 p_1^2},\quad A_1(p_1) =\frac{g_{\rm d}^4 m_\chi^2  \left(\cot ^2\left(\frac{\theta_{\rm min}}{2}\right)-2 \ln
   \left[\csc^2 \left(\frac{\theta_{\rm min}}{2}\right)\right]\right)}{2 p_1^2}.
\ee
We note that it is not necessary to compute here the coefficients with $l\geq2$ since these DR moments are suppressed due to the self-interactions of the non-Abelian gauge bosons ($\dot{\kappa}_{\rm DR-DR}\gg H$). The DM and DR opacities are then
\be
\dot{\kappa}_{\rm DR-DM} = -a\frac{5 \alpha_{\rm d}^2}{2 \pi} n_\chi^{(0)}  \frac{1}{T_{\rm DR}^2} \ln{\alpha_{\rm d}^{-1}},\qquad \dot{\kappa}_\chi = -a (N^2-1)\frac{\pi\alpha_{\rm d}^2}{9} \frac{T_{\rm DR}^2}{m_\chi}\ln{\alpha_{\rm d}^{-1}},
\ee
where $\alpha_{\rm d} \equiv g_{\rm d}^2/(4\pi)$, and where we have taken $\csc^2{(\theta_{\rm min}/2)}=\alpha_{\rm d}^{-1}$  \citep{Buen-Abad:2015ova}. The opacity coefficients are then
\be
a_0 = \frac{5 \alpha_{\rm d}^2}{2\pi} \frac{ \tilde{\rho}_{\rm crit}}{m_\chi \xi^2  T_{{\rm CMB},0}^2} \ln{\alpha_{\rm d}^{-1}},\qquad a_{n\geq1} = 0.
\ee
The ETHOS mapping is thus of the form
\be
\Big{\{}m_\chi, g_{\rm d}, \xi,\eta_\chi,\eta_{\rm DR} \Big{\}} \rightarrow \left\{\omega_{\rm DR},a_0\right\}.
\ee
\subsection{Application: Shape of the linear matter power spectrum}\label{sec:power_shape}
In previous sections, we have established that the shape and amplitude of the linear matter power spectrum of models where DM couples to a relativistic component can entirely be described in terms of a set of effective ETHOS parameters (in addition, of course, to the standard $\Lambda$CDM parameters). In this section, we illustrate the impact of different choices of these parameters on the linear matter power spectrum, focusing primarily on the combination $\{a_n\,, \alpha_l\}$. 
\begin{figure}[t]
\includegraphics[width=0.497\columnwidth]{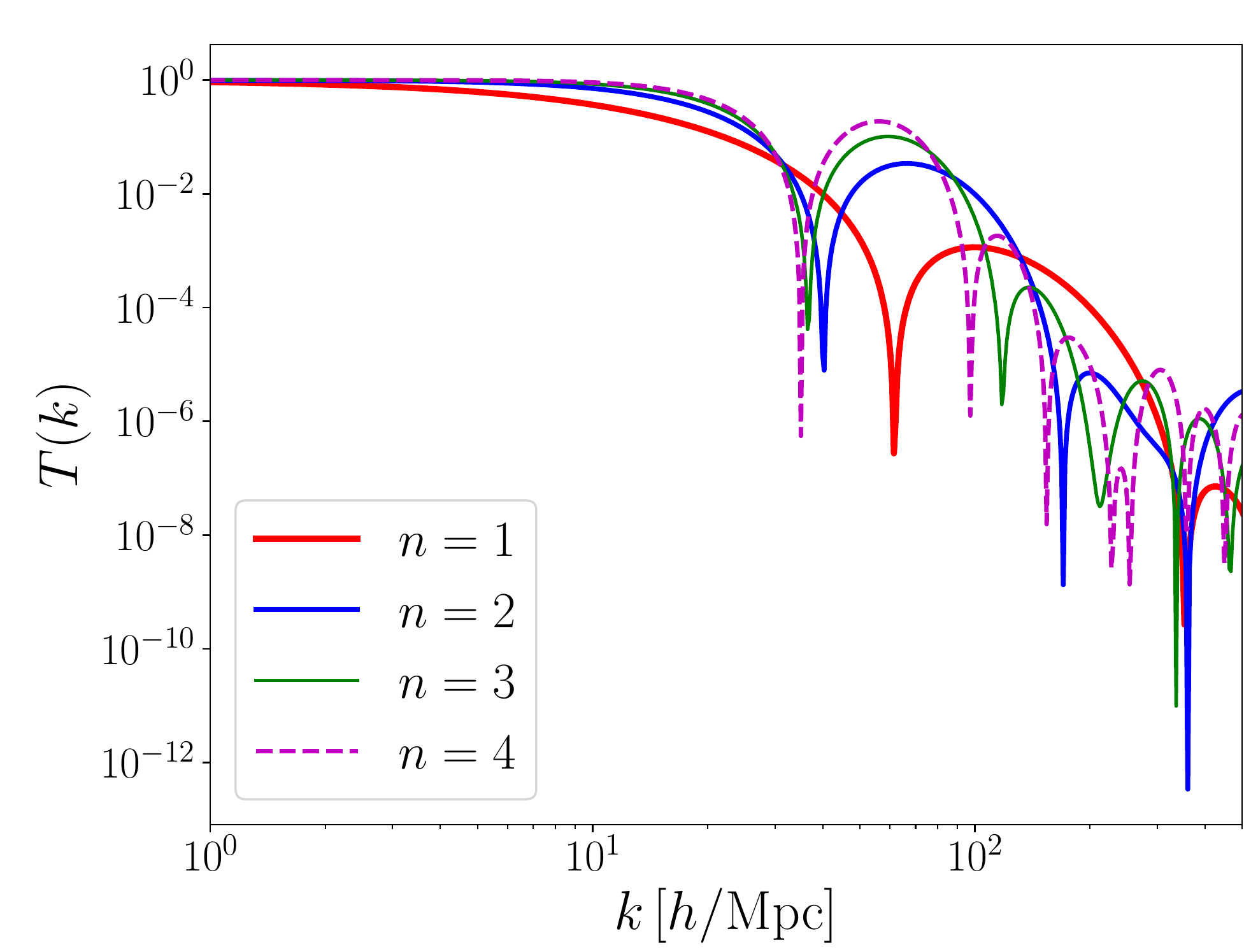}
\includegraphics[width=0.497\columnwidth]{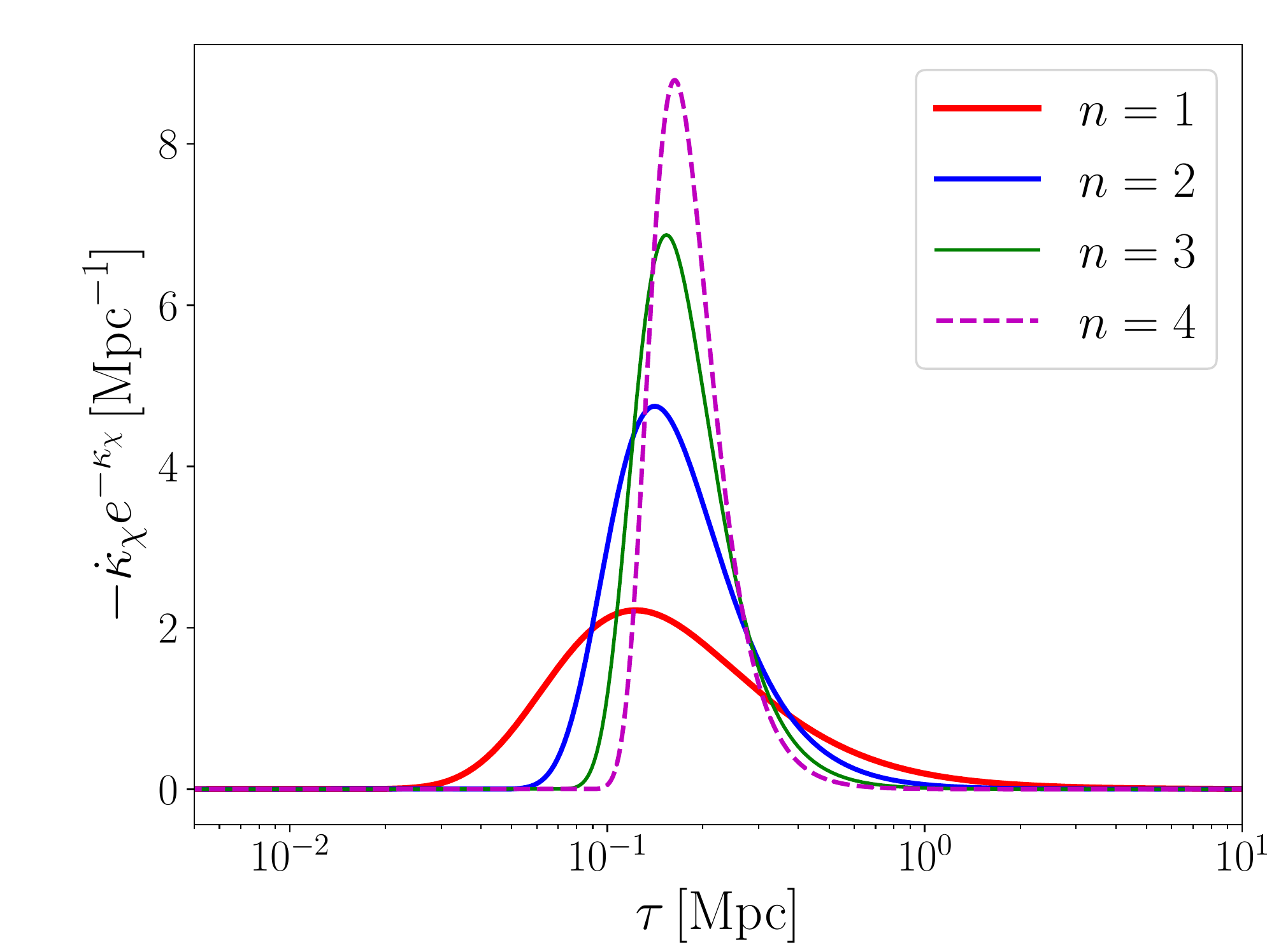}
\caption{\emph{Left panel}: Transfer function $T(k) \equiv P_{\rm ETHOS}(k)/P_{\rm CDM}(k)$ for four different exponents $n$ parametrizing the redshift dependence of the DM drag opacity $\dot{\kappa}_\chi= - (\Omega_{\rm DR}h^2) a_n(4/3)(1+z)^{n+1}/(1+z_{\rm D})^n$. The values of $a_n$ are chosen such that all models have the same DM drag epoch $z_{\rm drag}$, which we define via the criterion $-\dot{\kappa}_\chi(z_{\rm drag}) = \mathcal{H} (z_{\rm drag})$. The actual values used are $\{a_1,a_2,a_3,a_4\} = \{6.56,3.5\times10^{1}, 1.86\times10^{2}, 9.95\times10^{2}\}$ Mpc$^{-1}$. All models assume $\omega_{\rm DR} = 1.35\times10^{-6}$ , $\alpha_l = 1$, and $b_n = 0$. For completeness, we also used $\xi = 0.5$, $m_\chi = 10$ GeV, and $d_n =  a_n$, but the results shown above are insensitive to these specific choices. \emph{Right panel}: Dark matter drag visibility function for the same models as the left panel. The DM drag visibility function is essentially the probability distribution function for the time at which a DM particle last scatter off DR.\label{fig:Transfer_for_diff_n}}
\end{figure} 

The left panel of Fig.~\ref{fig:Transfer_for_diff_n} illustrates the matter transfer function $T(k) \equiv P_{\rm ETHOS}(k)/P_{\rm CDM}(k)$ for four different exponents $n$ parametrizing the redshift dependence of the DM and DR opacities. The models are normalized such that they all have the same DM drag epoch $z_{\rm drag}$ which we define via the criterion $-\dot{\kappa}_\chi(z_{\rm drag}) = \mathcal{H} (z_{\rm drag})$. All other parameters are kept fixed as indicated in the figure caption.  We observe that as $n$ is increased, the frequency of dark acoustic oscillations (DAO) increases and the transfer function begins departing from its CDM value at larger wave numbers (smaller scales). This is due to the faster decoupling time scale associated with larger values of $n$. We illustrate this in the right panel of Fig.~\ref{fig:Transfer_for_diff_n} where we display the DM drag visibility function $-\dot{\kappa}_\chi e^{-\kappa_\chi}$ for the same models as in the left panel. We observe that a larger value of the exponent $n$ corresponds to a narrower DM drag visibility function. Since $\dot{\kappa}_\chi/\mathcal{H}\propto (1+z)^n$, a larger value of $n$ indeed implies a faster transition from the tightly coupled regime $\dot{\kappa}_\chi/\mathcal{H}\gg1$ to the decoupled regime $\dot{\kappa}_\chi/\mathcal{H}\ll1$. In contrast, as $n$ approaches $0$, DM spends more time in the weakly coupled regime and a broader range of $k$-modes can be affected by the dark sector physics. This is particularly apparent for the $n=1$ model where a large range of $k$-modes are damped by DR diffusion. A longer period spent in the weak coupling regime also implies that the damping envelope significantly departs from the exponential relation $e^{-(k/k_{\rm damp})^2}$ derived in the tight-coupling limit \citep{Hu:1995en}.
\begin{figure}[b]
\includegraphics[width=0.497\columnwidth]{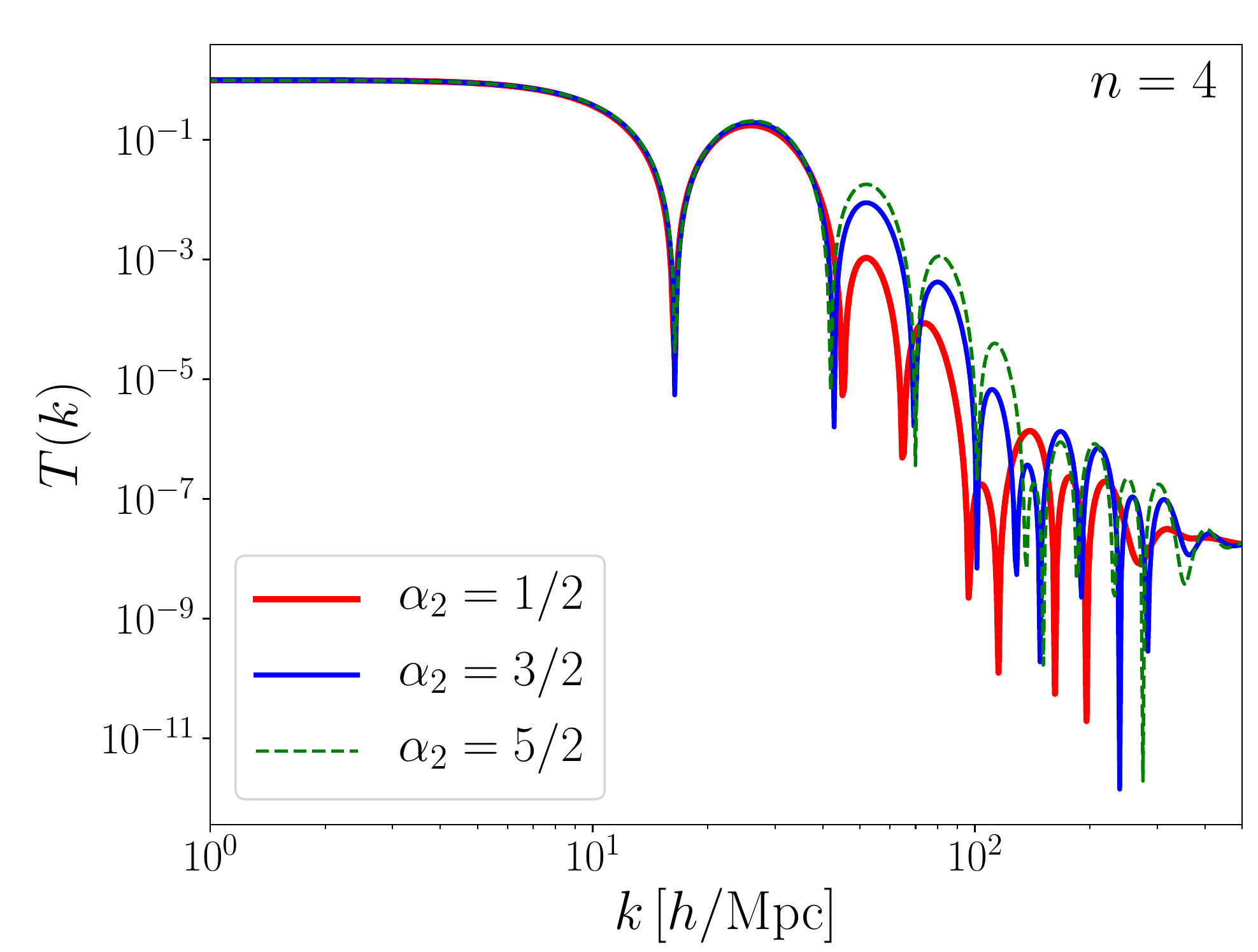}
\includegraphics[width=0.497\columnwidth]{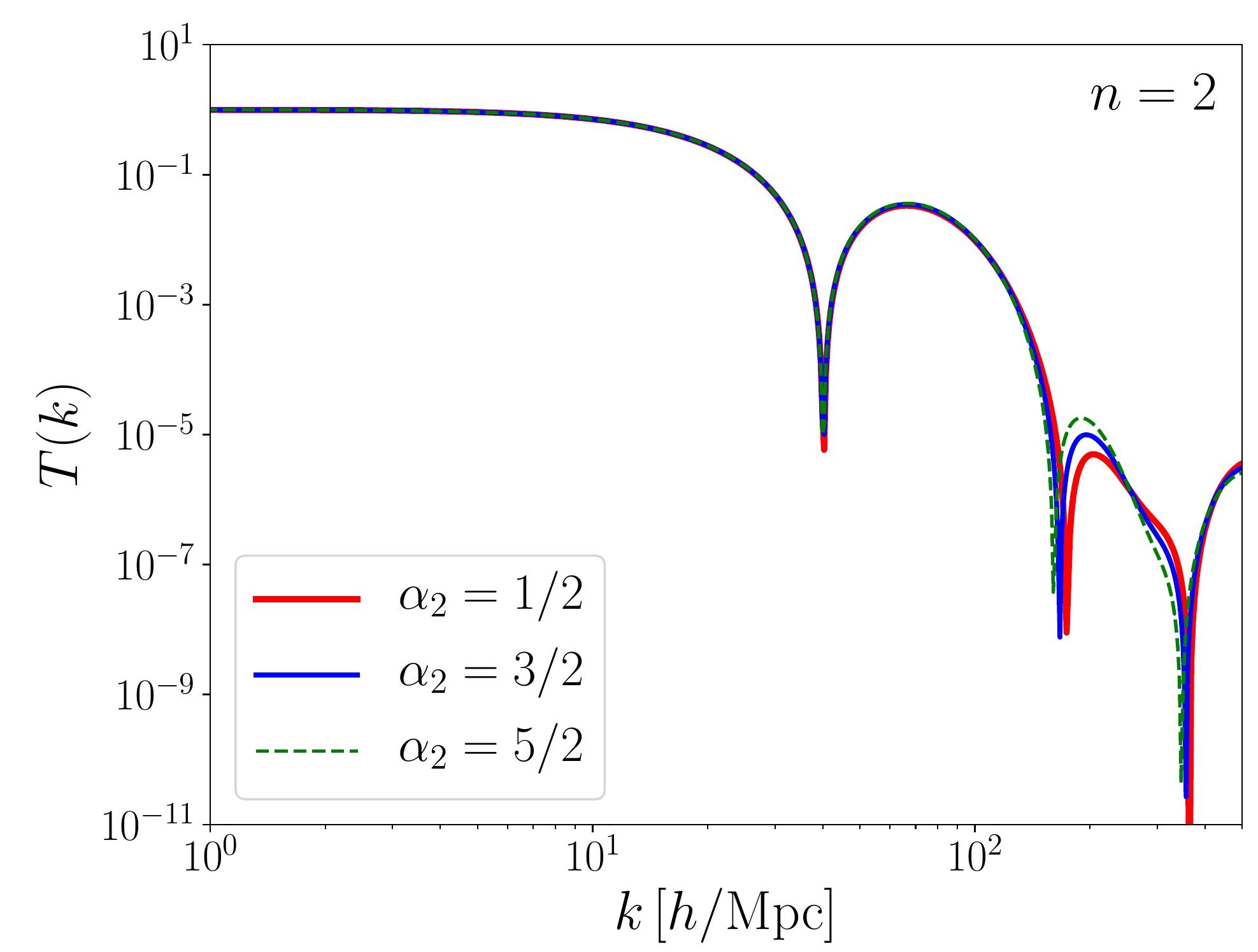}
\caption{\emph{Left panel}: Transfer function for three different values of $\alpha_2$ for a model characterized by a nonvanishing value of $a_4$. The model shown here assumes fermionic DR with $a_4 = 2.24\times10^{4}$ Mpc$^{-1}$, $\xi = 0.5$, $m_\chi = 2$ TeV, $\eta _{\rm DR} = \eta_\chi =2$, $b_n = 0$, and $\alpha_{l\geq3}=1$. \emph{Right panel}: Similar to the left panel but for a model with $a_2 = 3.5\times 10^1$ Mpc$^{-1}$. We assume fermionic DR with $\xi = 0.5$, $m_\chi = 10$ GeV, $\eta _{\rm DR} = \eta_\chi =2$, $b_n = 0$, and $\alpha_{l\geq3}=1$.  \label{fig:Transfer_for_diff_alpha_l}}
\end{figure} 

In Fig.~\ref{fig:Transfer_for_diff_alpha_l}, we study the impact of the angular coefficients $\alpha_2$ on the matter transfer function. Here, we choose models with a nonvanishing $a_4$ (left panel) and $a_2$ (right panel) coefficient, and vary the value of $\alpha_2$ from $1/2$ to $5/2$ while keeping everything else fixed. While we realize that it might not be possible to find a physical DM model realizing these different values of $\alpha_2$, our goal here is to illustrate the sensitivity of the DM distribution to these parameters. The left panel of Fig.~\ref{fig:Transfer_for_diff_alpha_l} shows that $\alpha_2$ has a significant effect on the damping tail of the matter transfer function, with a smaller value of $\alpha_2$ associated with more damping. We can understand this result by noting that the quantity $\alpha_2 \dot{\kappa}_{\rm DR-DM}$ controls the growth of the DR quadrupole which is associated with DR diffusion damping of DM perturbations. At a fixed value of the opacity $\dot{\kappa}_{\rm DR-DM}$, a smaller $\alpha_2$ leads to a faster growth of the DR quadrupole, which results in a stronger damping term. This can also be seen from the direct calculation of the Silk damping scale, which in the tightly coupled regime takes the approximate form
\be
r_{\rm SD}(\tau) \approx \pi \left( -\frac{1}{6}\int_0^\tau\frac{d\tau'}{\dot{\kappa}_\chi + \dot{\kappa}_{\rm DR-DM}}\left[  \frac{\dot{\kappa}_{\rm DR-DM}}{\dot{\kappa}_\chi + \dot{\kappa}_{\rm DR-DM}} + \frac{4\dot{\kappa}_\chi}{5(\alpha_2\dot{\kappa}_{\rm DR-DM} +\beta_2 \dot{\kappa}_{\rm DR-DR})}\right]\right)^{1/2}.
\ee
Thus, a larger value of $\alpha_2$ indeed corresponds to a smaller damping scale in the tightly coupled regime. One might ask whether this result holds in models that spend a significant amount of time in the weakly coupled regime. We illustrate this latter case in the right panel of Fig.~\ref{fig:Transfer_for_diff_alpha_l} where we display a model with a nonvanishing $a_2$ coefficient. There, we demonstrate that the matter transfer function is almost insensitive to $\alpha_2$. In these models, the broad DM drag visibility function effectively erases the memory of the specific value of $\alpha_2$, and the shape of the DM power spectrum is almost entirely dictated by $\dot{\kappa}_\chi$. This implies that a detailed calculation of the exact values of the angular coefficients is less important for models dominated by low-$n$ $a_n$ coefficients. 

\begin{figure}[t]
\includegraphics[width=0.497\columnwidth]{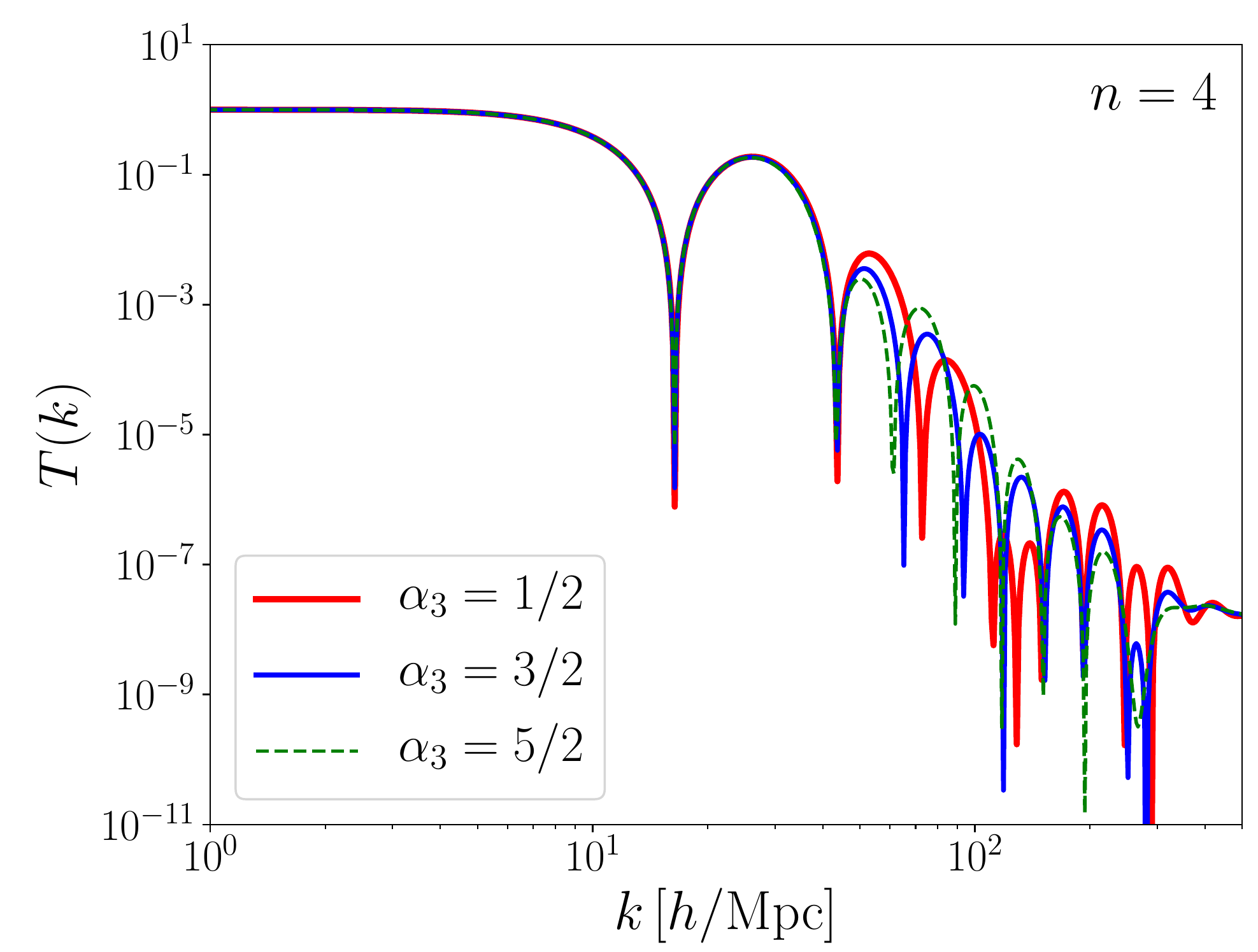}
\includegraphics[width=0.497\columnwidth]{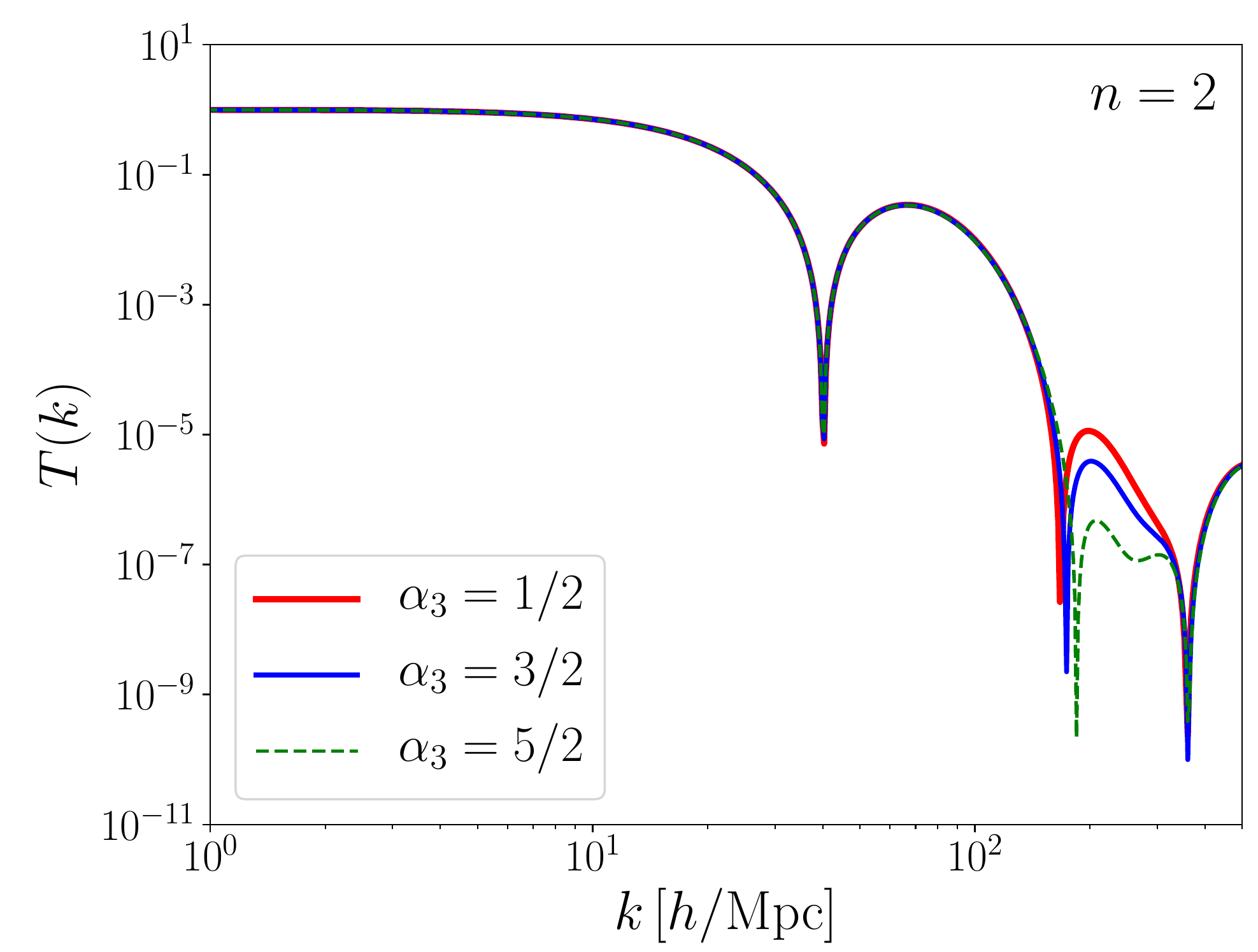}
\caption{\emph{Left panel}: Transfer function $T(k) \equiv P_{\rm ETHOS}(k)/P_{\rm CDM}(k)$ for three different values of $\alpha_3$ for a model characterized by a nonvanishing value of $a_4$. The model shown here assumes fermionic DR with $a_4 = 2.24\times10^{4}$ Mpc$^{-1}$, $\xi = 0.5$, $m_\chi = 2$ TeV, $\eta _{\rm DR} = \eta_\chi =2$, $b_n = 0$, $\alpha_2 =1$, and $\alpha_{l\geq4}=1$. \emph{Right panel}: Similar to the left panel but for a model with $a_2 = 3.5\times 10^1$ Mpc$^{-1}$ and $m_\chi = 10$ GeV.  \label{fig:Transfer_for_diff_alpha_3}}
\end{figure} 
In Fig.~\ref{fig:Transfer_for_diff_alpha_3}, we illustrate the impact of the next order angular coefficient $\alpha_3$. Similarly to Fig.~\ref{fig:Transfer_for_diff_alpha_l}, the left panel displays a model with a nonvanishing $a_4$ coefficient for three different choices of $\alpha_3$. We observe that this parameter does affect the shape of the damping envelope of the matter transfer function, but in a more intricate way than $\alpha_2$. Both the amplitude and phase of the second and subsequent acoustic oscillation peaks are affected by the value of $\alpha_3$, in contrast to $\alpha_2$ which mostly affected the amplitude of the damping envelope. In the right panel of Fig.~\ref{fig:Transfer_for_diff_alpha_3}, we illustrate the impact of $\alpha_3$ for a model characterized a nonzero value of $a_2$. As in the case of $\alpha_2$, the matter transfer function for $n=2$ displays little sensitivity to the angular coefficient $\alpha_3$. The second acoustic oscillation peak is marginally affected, but it is very unlikely that such a tiny feature has any effect on nonlinear structure formation. Again, the width of the DM drag visibility function for a model with low $n$ values tends to erase the memory of the angular dependence of the DM-DR scattering cross section. 

In summary, we have seen that for a fixed DM drag epoch, DM models characterized by opacities with weak redshift (or temperature) dependence generally display a broader drag visibility function, which tends to wash out the details of the angular dependence of the DM-DR scattering cross section. The wider visibility function also leads to a broader power spectrum damping envelope which assumes a different shape than the standard $e^{-k^2/k_{\rm damp}^2}$. On the other hand, DM models that have an opacity with a steep redshift dependence near the drag epoch are more sensitive to the details of the DM-DR scattering cross section encoded in the $\alpha_l$ coefficients. In general, as the redshift dependence of the opacity steepens, we expect the matter transfer function to display an increasing number of  essentially undamped DAOs. This so-called ``strong'' DAO regime \citep{Buckley:2014hja} in which the matter transfer function can undergo several undamped acoustic oscillations before being exponentially Silk damped occurs for instance in the atomic DM model \citep{Cyr-Racine:2013ab,Cyr-Racine:2013fsa}. There, the rapidly declining DM ionized fraction leads to an exponentially decreasing drag opacity near the DM drag epoch, hence yielding a matter power spectrum that is very sensitive to the details of the DM-DR interaction.
\medskip

In this section, we have provided a detailed exposition of the linear evolution of DM density perturbations that are coupled to some form of relativistic DR. We have computed how the angular dependence of the DM-DR scattering cross section enters the DR Boltzmann hierarchy and studied how the latter affects the shape of the linear matter power spectrum. We have also adopted a simple parametrization to describe the main physics determining the structure of the power spectrum. Within the ETHOS framework, this parametrization allows one to capture the full shape of the linear matter power spectrum with a small set of effective parameters, without the need to specify a particular DM particle physics model. This in turns enables a bottom-up view of structure formation in which we can use data to determine which DM physics is most relevant to observations, without the requirement of carrying out model-by-model analyses. Put in another way, by imposing observational constraints on the effective ETHOS parameters (such as $a_n$), we would be simultaneously restricting the available parameter space of all corresponding DM particle models, hence allowing for a much more general analysis.

%%%
\section{Dark matter self-interaction}\label{sec:self-interaction}
%%%
In the previous section, we have computed in detail the linear evolution of DM density 
fluctuations in the presence of new interactions with a relativistic component. In certain 
scenarios (see e.g.~\cite{Aarssen:2012fx,Bringmann:2013vra,Chu:2014lja,Buckley:2014hja}), the physics responsible for DM-DR interaction naturally leads also to significant DM self-scattering inside halos at later epochs of the Universe. Independently of this connection, allowed DM-DM interactions have also been considered as a way to alleviate some
of the potential shortcomings of the standard cold DM model at small scales (see our companion paper for a thorough review of these issues \citep{2016MNRAS.460.1399V}), and have been seen with a renewed interest 
due to the recent observation of an unaccounted-for displacement of the stars in a member 
galaxy of the cluster Abell 3827 relative to its gravitational center 
(\cite{Massey:2015dkw,Kahlhoefer:2015vua,Kahlhoefer:2013dca}). It is thus entirely natural to extend the ETHOS framework to allow for DM self-scattering. In this work, we focus exclusively on elastic DM scattering but we note that the ETHOS framework could be expanded to include dissipative DM interactions \citep{Randall:2014kta,Fan:2013yva,McCullough:2013jma,Fan:2013tia,Cyr-Racine:2013ab}.

DM self-interactions are usually quantified in terms of the 
momentum-transfer cross section over the DM mass,
\be
\frac{\sigma_T}{m_\chi} = \frac{1}{m_\chi}\int d\Omega \frac{d\sigma_{\chi\chi\rightarrow\chi\chi}}{d\Omega}(1-\cos{\theta}),
\ee
where $\theta$ is the scattering angle in the center-of-mass frame. This quantity
conveniently regulates divergences in the differential scattering cross section that appear 
for forward scattering, which are not relevant for our purposes since they do not change the 
DM distribution (see Ref.~\cite{Tulin:2013teo} for an extensive discussion).
In general, the transfer cross section is a velocity-dependent function 
\citep{Ackerman:2008gi,Feng:2009mn,ArkaniHamed:2008qn,2010PhRvD..81h3522B,Feng:2009hw,2011PhRvL.106q1302L,Hooper:2012cw,Tulin:2012wi,Tulin:2013teo} 
implying that astrophysical objects of different masses would be affected differently by 
DM self-scattering. This velocity dependence can naturally accommodate a scenario in which the transfer cross section is larger at small velocities, characteristic of dwarf galaxies, than at larger velocities, characteristic of galaxy clusters. In this way, models with DM self-scattering are a viable alternative to alleviate the dwarf-scale cold DM challenges, while at the same time avoid current constraints on the transfer cross section at cluster scales. For instance, Ref.~\cite{Kaplinghat:2015aga} finds that a value of 
$(\sigma_T/m_\chi)\sim 2$ cm$^2$/g is broadly consistent with observations of dwarf and 
low-surface brightness galaxies, while a lower value of $(\sigma_T/m_\chi)\sim0.1$ cm
$^2$/g is necessary to accommodate the observed density profile of galaxy clusters. A 
convenient way of parametrizing the DM self-interaction cross section is then to split 
$(\sigma_T/m_\chi)$ in a certain number of relative-velocity bins centered on the typical 
velocity dispersion of astrophysical objects that span the range of scales of interest for
studies of the DM distribution inside galaxies and galaxy clusters. Quantitatively, we 
implement this by averaging the DM transfer cross section over a Maxwellian distribution, 
\be\label{eq:ETHOS_vel_ave}
\frac{\langle \sigma_T \rangle_{v_{M}}}{m_\chi} = \int d^3 v \,  f(v;v_{ M})\,\frac{\sigma_T(v)}{m_\chi},\qquad\text{where}\qquad f(v;v_{ M}) = \frac{e^{- v^2/2 v_{ M}^2}}{(\sqrt{2\pi}v_{ M})^3},
\ee
where $v$ is the relative velocity of two colliding DM particles, and  
$v_{ M}$ is the most probable speed (of a single particle) for a halo of 
mass scale $M$ {\bf }. Following the same philosophy as for the matter power spectrum, we can define an ETHOS mapping between the DM particle properties and the DM transfer cross 
sections relevant to structure formation
\be\label{eq:SI_cross_sec_map_ETHOS}
\Big\{m_\chi,\{m_i\}, \{g_i\} \Big\} \rightarrow \Bigg\{\frac{\langle \sigma_T \rangle_{30}}{m_\chi},\frac{\langle \sigma_T \rangle_{220}}{m_\chi},\frac{\langle \sigma_T \rangle_{1000}}{m_\chi}\Bigg\},
\ee
where $m_i$ and $g_i$ are masses and coupling constants appearing in the DM particle 
theory, respectively. Here, we have chosen three relative-velocity bins centered at $v_{\rm rel} = 30$, $220$, and $1000$ km/s, which very roughly correspond to typical velocities encountered in dwarf galaxies, Milky-Way-size galaxies and galaxy clusters, respectively.  We note that the number of velocity bins and the choice of their central values is arbitrary, but we find that three or four values of $\langle \sigma_{T}\rangle_{v_{ M}}/m_\chi$ spread over a broad range of mass scales are generally sufficient to characterize a given model in terms of its main effects in three relevant regimes for structure formation: (i) the dwarf-scale regime where the cold DM model is being challenged, and where the transfer cross section is largely unconstrained, (ii) the intermediate-scale regime where a large cross section can lead to the evaporation of subhaloes in Milky-Way-size galaxies, and (iii) the cluster-scale regime where observations put the strongest constraints to the transfer cross section. We caution, however, that this mapping is likely an oversimplification of the actual self-interaction dynamics that is taking place inside a DM halo. It nevertheless provides clear guidelines on whether a given model can be compatible with observations at the several key velocities where measurements are available. More importantly, it enables the interpretation of simulation results that where obtained for one particular DM model in terms of  other DM theories. For instance, if two models have identical values of $\langle \sigma_T \rangle_{30}$, simulations of isolated dwarfs in each model are likely to yield similar results, even though their transfer cross section may differ significantly at $v=1000$ km/s.

While one may in principle take a very phenomenology driven spirit and allow for any 
velocity dependence of $\sigma_T$ for this kind of mapping, there are a few 
cases that are particularly interesting from the point of view of realistic model building. We discuss these cases below from the point of view of the ETHOS mapping suggested in Eq.~\eqref{eq:SI_cross_sec_map_ETHOS}.
\subsection{Constant self-interaction cross section}\label{sec:constant_cross_sec}
We first consider the simplest case of a constant, velocity-independent self-interaction cross section, which corresponds to the original 
proposal as put forward by Ref.~\cite{Spergel:1999mh} and may be thought of as arising
from a pointlike, effective interaction mediated by some heavy messenger. In order for 
that limit to apply, however, the messenger should be much heavier than the DM particle,
which makes it somewhat challenging to construct concrete models with a self-interaction 
rate that is large enough to visibly deviate from the predictions of cold DM at the scale of dwarf galaxies. From the perspective of ETHOS, these models are trivial since they only depend on a single parameter
\be
\Big\{m_\chi,\{m_i\}, \{g_i\} \Big\} \rightarrow \Bigg\{ \frac{\sigma_T}{m_\chi}\Bigg\}.
\ee
We also note that such constant cross section models, contrary to velocity-dependent models, face difficulties in satisfying observational constraints if invoked as a solution to the dwarf-scale challenges to the cold DM model: the observed ellipticity of clusters 
\citep{Yoshida:2000uw}, the survivability of large galaxies in clusters or dwarf galaxies in 
the  Local Group \citep{Gnedin:2000ea}, the imminent relaxation of halo cores 
to even denser states in a ``gravothermal catastrophe'' \citep{Balberg:2002ue}, and the limit on DM collisionality from the mergers of galaxy cluster \citep{Randall:2007ph}. 
\subsection{Yukawa-type self-interaction}
The second well-motivated example we consider here has a velocity dependence dictated by a Yukawa potential between the DM particles \cite{2011PhRvL.106q1302L,Tulin:2012wi,Tulin:2013teo}. This may not only 
alleviate the constant cross-section problems mentioned in Sec.~\ref{sec:constant_cross_sec} but, in fact,  
potentially address all shortcomings of cold DM simultaneously by invoking the 
same physics in explaining DM-DM and DM-DR interactions \citep{Aarssen:2012fx,Bringmann:2013vra,Chu:2014lja}. From a microscopic point of view, such an interaction is extremely well motivated and 
corresponds to the existence of a new light messenger particle $\phi$ that mediates this 
``dark force''.
Assuming a coupling constant $g_\chi$ in the interaction term between the DM particles 
and the (vector or scalar) messenger $\phi$ in the Lagrangian, the resulting Yukawa 
potential is given by
\be
  V(r)=\pm \frac{\alpha_\chi}{r}e^{-m_\phi r}\,,
\ee
where $\alpha_\chi\equiv g_\chi^2/(4\pi)$. For scalar $\phi$ as well as self-conjugate DM, 
like Majorana fermions, the potential is always repulsive (+); otherwise it can be both 
attractive ($-$) and repulsive. For Dirac DM coupling to vector particles, for example, scattering events take place equally often for both types of potentials and one thus has to compute the {\it average} of the 
resulting scattering cross sections. We note that in order to compute the full velocity dependence of $\sigma_T$ in 
the presence of a Yukawa potential, one needs to perform a partial wave analysis of the 
Schr\"odinger equation (see, e.g., \cite{vandenAarssen:2012ag} and in particular 
\cite{Tulin:2013teo} for a very nice technical description). In the classical limit ($m_\chi v\gg m_\phi$), convenient 
parameterizations for the transfer cross section can be obtained by considering the 
analogous situation of screened Coulomb scattering in a plasma\footnote{Note that these 
expressions improve those used earlier 
\citep{khrapak1,khrapak2,khrapak3,Feng:2009hw,Tulin:2012wi}.  In particular,  the 
parametrization for intermediate values of $\beta$ (where $\sigma_T v$ takes its maximum 
for the range of parameters we are interested in here) now connects smoothly to the 
strong interaction (large $\beta$) and Coulomb (small $\beta$)  limits of the scattering 
cross section, and is closer to the actual numerical results obtained by Refs.
\cite{khrapak1,khrapak2,khrapak3}.}:
\ba
\label{sigT-}
 \sigma_T^-&=&\begin{cases}
          \frac{2\pi}{m_\phi^2}\beta^2\ln(1+\beta^{-2}) & \beta\lesssim10^{-2}\\
          \frac{7\pi}{m_\phi^2}\frac{\beta^{1.8}+280(\beta/10)^{10.3}}{1+1.4\beta+0.006\beta^4+160(\beta/10)^{10}} & 10^{-2}\lesssim\beta\lesssim10^2\\
          \frac{0.81\pi}{m_\phi^2}(1+\ln\beta-(2\ln\beta)^{-1})^2 & \beta\gtrsim10^2
         \end{cases}\\
\label{sigT+}
 \sigma_T^+&=&\begin{cases}
          \frac{2\pi}{m_\phi^2}\beta^2\ln(1+\beta^{-2}) & \beta\lesssim10^{-2}\\
          \frac{8\pi}{m_\phi^2}\frac{\beta^{1.8}}{1+5\beta^{0.9}+0.85\beta^{1.6}} & 10^{-2}\lesssim\beta\lesssim10^4\\
          \frac{\pi}{m_\phi^2}(\ln2\beta-\ln\ln2\beta)^2 & \beta\gtrsim10^4
         \end{cases}
\ea
where $\beta\equiv2\alpha_\chi m_\phi/(m_\chi v^2)$. 
In both cases, the momentum-weighted cross section $\sigma_T v$ rises 
moderately with $v$ until it peaks at $v=v_{\rm max}$, and then falls off sharply\footnote{We note that our improved parametrization of the transfer cross sections yields slightly different values of $v_{\rm max}$ and $\sigma_T^{\rm max}\equiv\sigma_T(v_{\rm max})$ than those appearing in the literature.  For a purely repulsive 
 potential, Eq.~(\ref{sigT+}) gives $v^2_{\rm max}=0.1\alpha_\chi m_\phi/m_\chi$ and $\sigma_T^{\rm max}=31.2 m_\phi^{-2}$. In cases where one needs to take the 
 average of Eqns.~(\ref{sigT-}) and (\ref{sigT+}), we find 
 $v^2_{\rm max}=0.52\,\alpha_\chi m_\phi/m_\chi$ and $\sigma_T^{\rm max}=21.8\, m_\phi^{-2}$.}. This 
phenomenologically allows large cross sections on dwarf galaxy scales, $v\sim10\!\!-\!\!50$\,km/s, 
while having much smaller cross sections at cluster scales, with $v\sim1000$\,km/s. 
While the above parametrizations agree extremely well 
with the full analytical results in the classical regime ($m_\chi v\gg m_\phi$), we caution that they fail to reproduce 
the resonances appearing for an attractive potential at slightly larger mediator masses, 
$m_\phi\gtrsim m_\chi v$, and also do not show the correct behavior in the (nonresonant) 
Born regime ($m_\phi\gg\alpha_\chi m_\chi$). We refer the reader to the Appendix of Ref.~\cite{Tulin:2013teo} for a discussion of these other regimes.

\begin{figure}[t]
\includegraphics[width=0.497\columnwidth]{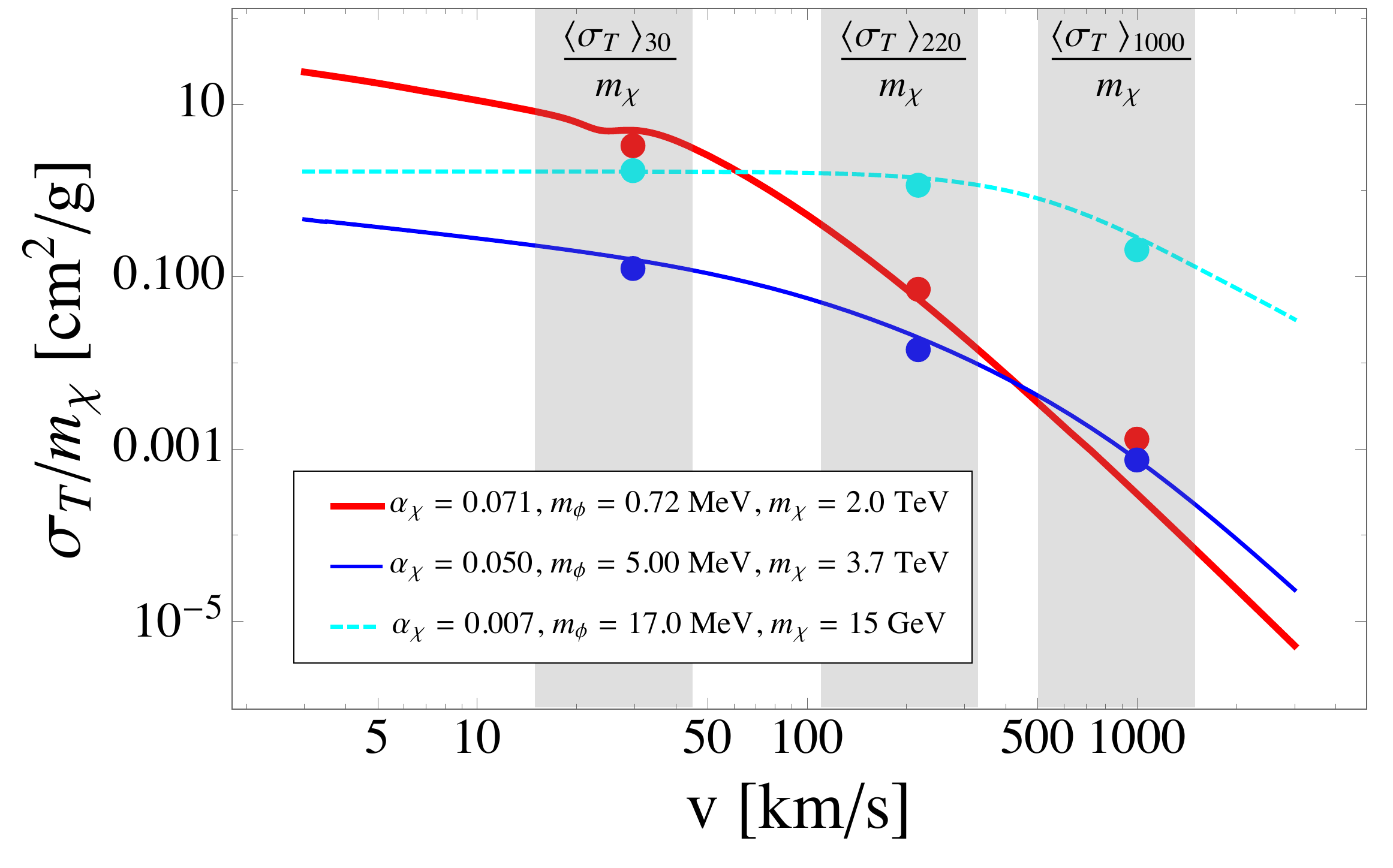}
\includegraphics[width=0.497\columnwidth]{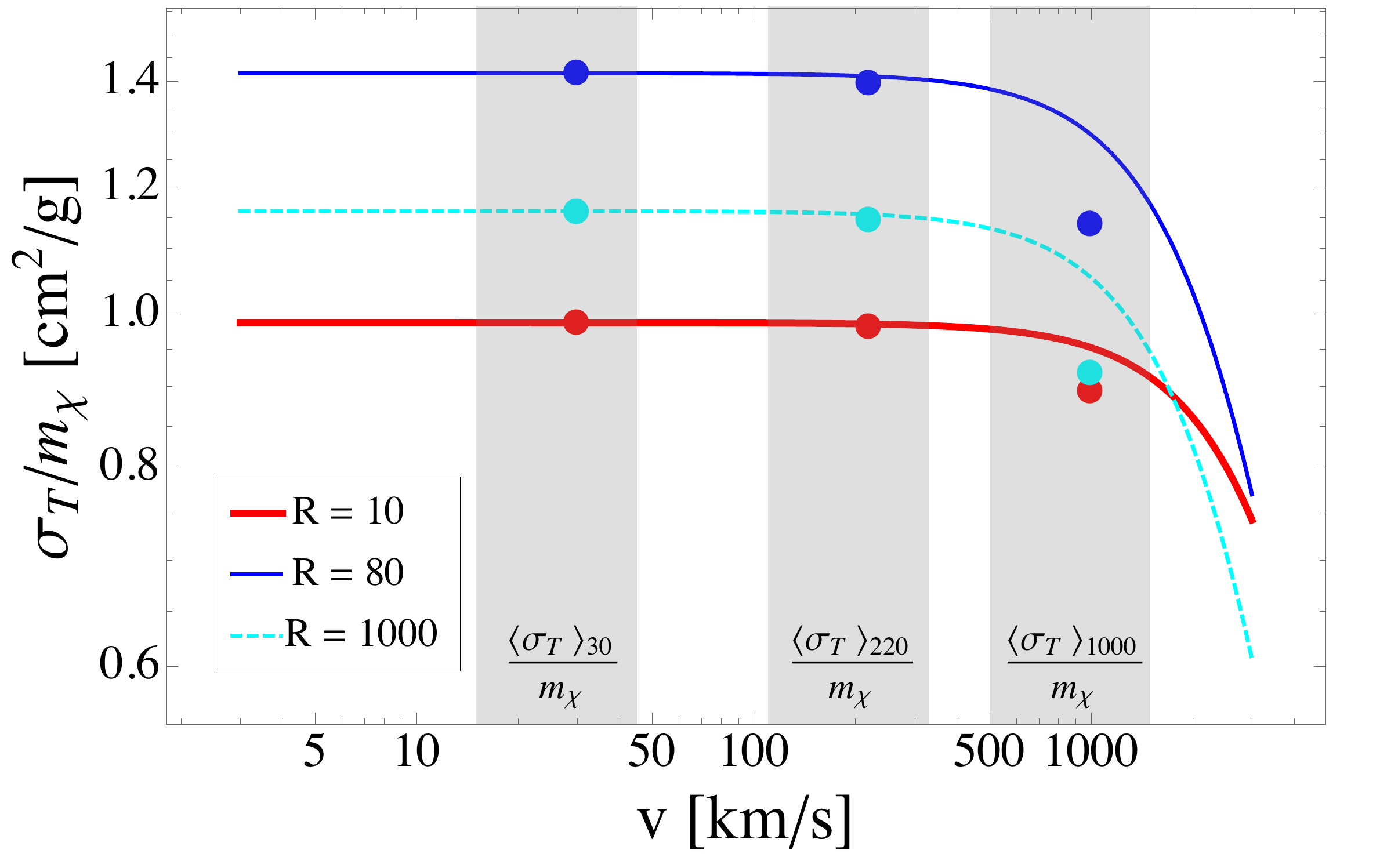}
\caption{\emph{Left panel}: Velocity dependence of the self-interaction cross section over mass  for DM interacting via a Yukawa potential mediated by a messenger particle $\phi$ \cite{2011PhRvL.106q1302L,Hooper:2012cw,Tulin:2012wi,Tulin:2013teo}. The model shown with the thick red solid curve is an example of a symmetric DM model that primarily scatters in the classical regime ($m_\chi v\gg m_\phi$) with a momentum-transfer cross sections given by the average of Eqs.~\eqref{sigT-} and \eqref{sigT+}. The thin solid blue line is an example of asymmetric DM that primarily scatters in the classical regime with a momentum-transfer cross sections given by Eq.~\eqref{sigT+}. The dashed cyan curve is an example of an asymmetric DM model similar to the model put forward in Ref.~\citep{Kaplinghat:2015aga}. This model primarily scatters in the nonperturbative regime ($m_\chi v\lesssim m_\phi$) and we refer the reader to the Appendix of Ref.~\citep{Tulin:2013teo} for an explicit analytical formula that is valid in this regime. In all cases, the colored points show the average values $\langle \sigma_{T}\rangle_{v_{M}}/m_\chi$ (as defined in Eq.~\eqref{eq:ETHOS_vel_ave}) for the three typical velocity ranges shown here by the gray bands. Note that the width of the gray bands is for illustration purposes only.   \emph{Right panel}: Similar to the left panel but for atomic DM models \citep{Goldberg:1986nk,Kaplan:2009de,Behbahani:2010xa,Kaplan:2011yj,Cline:2012is,Cyr-Racine:2013ab,Cline:2013zca,Cline:2013pca}. Here, the models are labeled by the value of $R$, which is the mass ratio of the two particles forming the dark atom. We show the approximate fitting formula for the momentum-transfer cross section given in Eq.~(10) of Ref.~\cite{Cline:2013pca} with a dark fine-structure constant value of $\alpha_D = 0.05$. For all the cases shown, the DM mass is determined from the relation $m_\chi = (R/\alpha_D)^{2/3}$ GeV \cite{Cline:2013zca}. The colored points show the values of  $\langle \sigma_{T}\rangle_{v_{ M}}/m_\chi$ for each typical velocities $v_{M}$.\label{fig:SI_cross_section_ETHOS}}
\end{figure}

We illustrate in the left panel of Fig.~\ref{fig:SI_cross_section_ETHOS} three different DM models that can interact via a Yukawa potential. The curves show the full velocity dependence of the models, while the points display the ETHOS effective values of the transfer cross section over mass, according to the map given in Eq.~\eqref{eq:SI_cross_sec_map_ETHOS}.  Note that the width of the different velocity bins is somewhat arbitrary and chosen only for illustration purposes. The models plotted here clearly show the diversity of velocity dependence that is possible in this type of models. For instance, the models represented by the solid red line and the dashed cyan line have similar transfer cross sections at $v=30$ km/s while having very different cross sections at $v = 1000$ km/s. Conversely, the models given by the solid blue and red lines have similar transfer cross sections at $v=1000$ km/s, while their cross sections are essentially 2 orders of magnitude apart at $v=30$ km/s. We also see in the left panel of Fig.~\ref{fig:SI_cross_section_ETHOS} that representing the entire velocity dependence of the transfer cross section by three effective values captures the essential behavior. Even if one was to naively interpolate between the given pivot points and use this in a simulation, it is very likely that this would yield a very similar structure formation scenario.

\subsection{Atomic dark matter}
The last example we consider here is atomic DM \cite{Goldberg:1986nk,Kaplan:2009de,Behbahani:2010xa,Kaplan:2011yj,Cline:2012is,Cyr-Racine:2013ab,Cline:2013zca,Cline:2013pca,Choquette:2015mca}, which is a composite model where two oppositely charged particles form a stable bound state. For simplicity, we restrict ourselves here to the case of dark atoms held together by an unbroken $U(1)$ force. In this case, the formation of dark atoms through a recombinationlike process in the early Universe requires the dark fine-structure constant $\alpha_D$ to be large enough ($\alpha_D\gtrsim 0.01$) \cite{Cyr-Racine:2013ab}. If this bound is satisfied, most of the DM forms neutral bound states and the long-range force mediated by the dark $U(1)$ interaction is efficiently screened. The elastic scattering cross section of dark atoms has a very rich structure due to the multiple resonances that appear at different collision energies \cite{Cline:2013pca}.  At low velocities however, the momentum-transfer cross section is approximately constant with a typical size given by $\sigma_T\sim 100 a_{\rm Bohr}'^2$, where $a_{\rm Bohr}'$ is the dark atom's Bohr radius. At large velocities, the transfer cross section takes a Coulomb-like form with $\sigma_T\propto 1/v^4$. To be quantitative, we adopt the fitting formula provided in Ref.~\cite{Cline:2013pca}
\be
\sigma_T \approx \frac{a_{\rm Bohr}'^2}{A_0(R)+A_1(R) (E/\epsilon_D)+ A_2(R) (E/\epsilon_D)^2},
\ee
where $A_0$, $A_1$, and $A_2$ are dimensionless function that depend on the mass ratio $R$ of the two constituent particles of the dark atom, and where $\epsilon_D \equiv \alpha_D^2 \mu_D$ is the dark atom Rydberg constant ($\mu_D$ is the reduced mass of the dark atom's constituents). 

We illustrate in the right panel of Fig.~\ref{fig:SI_cross_section_ETHOS} the momentum-transfer cross section over mass for three different dark atom models. We label the models by their value of the mass ratio $R$; the values of the other relevant parameters are given in the figure caption. As before, the colored points show the effective ETHOS values of the transfer cross section over mass for the mapping given in Eq.~\eqref{eq:SI_cross_sec_map_ETHOS}. For the three models shown here, we observe that the velocity dependence is very mild over the range of velocities relevant to a broad spectrum of astrophysical objects (note for instance the difference of the y-axis between the left and right panels of Fig.~\ref{fig:SI_cross_section_ETHOS}). However, the qualitative behavior of dark atom scattering is similar to the nonperturbative scattering limit ($m_\chi v\lesssim m_\phi$) of the Yukawa DM model presented in the previous subsection.\footnote{See the dashed cyan line of the left panel of Fig.~\ref{fig:SI_cross_section_ETHOS}} This reinforces the idea that the ETHOS framework can encompass multiple models using a simple parametrization.

%%%
\section{ETHOS: Mapping particle models to structure formation scenarios}\label{sec:ETHOS}
%%%
In the standard cold DM paradigm, DM is assumed to be nonrelativistic and to interact primarily via the gravitational force. These simple hypotheses have been extremely successful at explaining the structure of the Universe on large scales. However, we must keep in mind that this success does not necessarily preclude the existence of nontrivial DM microphysics that could affect structure formation at smaller scales, where these hypotheses remain untested. Indeed, causality dictates that new nongravitational interactions in the DM sector can only modify the matter distribution on small scales, leaving large scales intact. Many models have been proposed that either allow for DM self-interactions inside halos at late times, or for interactions between DM and other particles in the early Universe, or both (see Sec.~\ref{Introduction} and references therein). An immediate difficulty in exploring these models is that structure formation on small scales is highly nonlinear, requiring expensive high-resolution simulations in order to make clear predictions that can be compared with observations. The cost of these simulations renders nearly impossible the task of a systematic exploration of all DM models that lead to modified small-scale structures. To address this situation, we develop here an ``Effective THeory Of Structure formation" (ETHOS), in which the DM microphysics is systematically mapped to effective parameters that directly control astrophysical structure formation. These effective parameters fully describe the linear evolution of the growth of structures and provide a convenient parametrization for DM self-interactions. These two ingredients can then serve as the input for simulations to follow the growth of structures in the nonlinear regime. The advantage of developing ETHOS is clear: {\it all DM particle models that map to a given effective ETHOS model can be constrained at the same time by comparing a single simulation of the effective ETHOS model with observations at no extra computational cost}. 

In Appendix \ref{sec:Gen_Boltz_eqs} (and summarized in Sec.~\ref{sec:linear_mat_power}), we have performed a detailed analysis of the Boltzmann equation governing the evolution of DM  (including DM dark radiation interactions and DR self-interactions), and have determined that the structure of the linear matter power spectrum can be \emph{entirely} determined (up to second-order effects) by a set of opacity and angular coefficients given by
\be
\Big\{\omega_{\rm DR}, \{a_n,\alpha_l\},\{b_n,\beta_l\}, \{d_n, m_\chi,\xi\}\Big\}.
\ee
Moreover, we have seen that, to a good approximation, the subset $\big\{\omega_{\rm DR},\{a_n,\alpha_l\}\big\}$ is largely responsible for setting the broad structure of the linear matter power spectrum, with the other parameters providing relatively small corrections.  The set of $l$-dependent coefficients $\alpha_l$ encompasses information about the angular dependence of the DM-DR scattering cross section, whereas $a_n$ are the coefficients of the power-law expansion in temperature (redshift) of the DM drag opacity caused by the DM-DR interaction. In Sec.~\ref{sec:self-interaction}, we have introduced a simple parametrization for the DM self-interaction cross section based on averages of the transfer cross section evaluated at a few velocities $v_{M}$ relevant to key astrophysical objects (dwarf galaxies, Milky-Way-size galaxies, and galaxy clusters). Taken together, the effective parameters describing a given ETHOS model are then
\be
\Xi_{\rm ETHOS} = \Bigg\{\omega_{\rm DR},\{a_n,\alpha_l\},\Big\{\frac{\langle \sigma_T \rangle_{v_{M_i}}}{m_\chi}\Big\}\Bigg\},
\ee
where we have allowed an arbitrary number of velocity reference points $v_{M_i}$. From the perspective of the structure formation theory, two models having identical effective parameters in ETHOS would yield universes populated by statistically identical DM structures. The above ETHOS parametrization thus allows the classification of DM theories with respect to their \emph{structure formation} properties, instead of their intrinsic particle properties. One might object that the mapping between particle theories and ETHOS scenarios is never exact since distinct DM models will always make slightly different predictions. However, the nonlinear nature of the evolution of small-scale structures is very effective at erasing the memory of small differences in the linear power spectrum \citep{Boehm:2003xr, 2016MNRAS.460.1399V}, hence making the mapping quite effective at classifying DM models in broad categories.

As a first application of the ETHOS framework, we present in a companion paper  \citep{2016MNRAS.460.1399V} high-resolution simulations of a few ETHOS models characterized by nonvanishing values of $a_4$ and $\alpha_{l\geq2}=3/2$, corresponding to the particle physics model described in Sec.~\ref{ref_model} (a massive DM particle interacting with a massless neutrino-like fermion via a new massive mediator). This application has the objective of using ETHOS to address at least two of the main challenges of the cold DM model regarding the DM distribution in the Milky Way, namely the missing satellite problem and the too big to fail problem. We stress however, that the scope of ETHOS goes beyond the cold DM challenges. It is a framework that generalizes structure formation to include viable DM phenomenology, offering a new and powerful tool to explore new DM physics.

%%%
\section{Conclusion}\label{sec:conclusions}
%%%

In this work we have described  an effective theory of structure formation (ETHOS), a framework that makes it possible to compute cosmological structure formation in a wide range of models in which nongravitational dark matter physics can have important effects on galactic and subgalactic scales. Within the ETHOS framework, dark matter models can be classified according to a small set of parameters describing their structure formation properties rather than their intrinsic particle properties.  This allows nonlinear structure formation to be studied in a model-independent and computationally efficient fashion.   Rather than running different structure formation simulations to explore the parameter spaces of individual particle models, simulations that cover phenomenologically interesting regions of the ETHOS parameter space can be used to simultaneously explore many microphysical models of dark matter physics.

Starting from the general Boltzmann equations describing the evolution of the dark matter and dark radiation phase-space densities, we have determined a standard procedure for mapping the detailed microphysics of particle dark matter models into a set of parameters that define the form of the linear power spectrum of matter density perturbations.  We have also described a similar mapping from microphysics to an astrophysically motivated parametrization of the dark matter self-interaction cross section that captures the main effects of self-interactions on dark matter halos at different mass scales.   
Taken together, these ETHOS parameters fully describe the dark matter physics required to simulate cosmological structure formation and we have explicitly demonstrated this procedure by giving several examples of well-motivated particle models that have been discussed in the literature.  

We note that as nonlinear evolution of small-scale structures is effective at erasing the memory of small differences in the linear power spectrum our parametrization may be more broadly applicable to dark matter physics beyond the types we discuss in detail here.   For instance, while the current ETHOS implementation focuses on nonrelativistic dark matter models interacting with a relativistic species it would be natural to extend this framework to include models where dark matter is warm rather than cold.  We note, however, that the current framework can already approximately capture the physics of warm dark matter at the level of producing an equivalent suppression scale in the linear power spectrum, and indeed when simulated leads to a nonlinear power spectrum nearly indistinguishable from a warm dark matter case \cite{2016MNRAS.460.1399V}.  We leave extensions of the formalism to other dark matter physics and a precise characterization of these nonlinear mappings to future work.

%%%%
\acknowledgements
%%%%
We thank Manoj Kaplinghat for useful suggestions, and Manuel A. Buen-Abad for pointing out an inconsistency with our definitions of perturbation variables in an earlier version of the manuscript. F.-Y. C.-R. acknowledges the support of the National Aeronautical and Space Administration ATP Grant No.~14-ATP14-0018 at Harvard University. K. S. gratefully acknowledges support from the Friends of the Institute for Advanced Study.  The research of K. S. is supported in part by a Natural Science and Engineering Research Council (NSERC) of Canada Discovery Grant.
The work of F.-Y. C.-R. was performed in part at the California Institute of Technology for the Keck Institute for Space Studies, which is funded by the W. M. Keck Foundation.  M. V. acknowledges support through a Research Support Committee (Reed Fund) award at the Massachusetts Institute of Technology. The Dark Cosmology Centre is funded by the Danish National Research Foundation. J. Z. is supported by the European Union under a Marie Curie International Incoming Fellowship, Contract No.~PIIF-GA-2013-62772. C.{\ }P. gratefully acknowledges the support of the Klaus Tschira Foundation.
%%%%
\appendix
%%%%
%
\section{The Collisional Boltzmann Equation for Dark Matter and Dark Radiation}\label{sec:Gen_Boltz_eqs}
In this Appendix, we present detailed derivations of the results given in Sec.~\ref{sec:Gen_Boltz_eqs_sum} above. The structure of this Appendix is as follows. We begin by studying in Sec.~\ref{app:sec:generalities} the structure of the Boltzmann equation dictating the evolution of dark matter in the early epochs of the Universe. We then study in Secs.~\ref{sec:coll_int} and \ref{sec:DRDR_coll} how the momentum and angular dependence of the physics responsible for the new interactions determine the structure of the collision integrals. In Secs.~\ref{sec:DR_equations} and \ref{sec:DM_eqs}, we use this latter structure to determine the final form of the cosmological perturbations equations for DM that couples to a relativistic species.
\subsection{Generalities and Setup}\label{app:sec:generalities}
We consider a scenario in which a single species of dark matter (DM, denoted by $\chi$) can interact with a relativistic component (denoted by $\tilde{\gamma}$) which we will generally refer to as dark radiation (DR)  Our goal is to determine the evolution of the DM and DR distribution functions, denoted by $\fD(\xx,{\bf P},\tau)$ and $\fR(\xx,{\bf P},\tau)$, respectively. Here, ${\bf P}$ is the canonical conjugate variable to $\xx$. We consider the situation where the only relevant process for DM is its 2-to-2 scattering with DR, $\chi\tilde{\gamma}\leftrightarrow\chi\tilde{\gamma}$, but allow for DR self-interactions through the process $\tilde{\gamma}\tilde{\gamma}\leftrightarrow\tilde{\gamma}\tilde{\gamma}$. We assume that the DM relic abundance is fixed at some high temperature (through e.g. thermal freeze-out) and we therefore neglect the effect of DM annihilation or decay on the evolution of $\fD(\xx,{\bf P},\tau)$. The evolution of the distribution functions is determined by the two coupled Boltzmann equations
\be
\frac{d\fD}{d\lambda} = C_{\chi\tilde{\gamma}\leftrightarrow\chi\tilde{\gamma}}[\fD,\fR],\qquad
\frac{d\fR}{d\lambda} = C_{\chi\tilde{\gamma}\leftrightarrow\chi\tilde{\gamma}}[\fR,\fD]+C_{\tilde{\gamma}\tilde{\gamma}\leftrightarrow\tilde{\gamma}\tilde{\gamma}}[\fR],
\ee
where $\lambda$ is an affine parameter that describes the trajectory of the observer and the right-hand sides of these equations are the collision terms defined with respect to $\lambda$ . In the conformal Newtonian gauge, the space-time metric takes the form
\be\label{metric}
ds^2 = a^2(\tau)[-(1+2\psi) d\tau^2+(1-2\phi) d\vec{x}^2],
\ee
where $a$ is the cosmological scale factor, $\tau$ is the conformal time, and $\phi$ and $\psi$ are the two gravitational potentials. We can choose to define the affine parameter in terms of the four-momentum $P$ of an observer $P^\mu \equiv \frac{d x^\mu}{d\lambda}$, where $x^\mu=(\tau,\vec{x})$ is a four-vector parametrizing the trajectory of the observer.  We note that this implicitly sets the affine parameter to be the proper time $\tau$ and selects a physically natural definition for the collision terms.  Using Eq.~(\ref{metric}), we can then write
 \be\label{eq:lambda}
 \frac{d}{d\lambda} = \frac{d\tau}{d\lambda}\frac{d}{d\tau}=P^0\frac{d}{d\tau}=\frac{E(1-\psi)}{a}\frac{d}{d\tau},
 \ee
 where we have used the dispersion relation $g_{\mu\nu} P^\mu P^\nu = -m^2$ and we have defined $E = \sqrt{p^2+m^2}$, $p= |\p|$, and $p^2 = g_{ij}P^iP^j$. We note that Eq.~(\ref{eq:lambda}) is valid to first order in perturbation theory. The left-hand side of the Boltzmann equation reads \citep{Dodelson-Cosmology-2003} 
 \be\label{eq:LHS_Boltzmann_eq}
 \frac{df}{d\tau} =\frac{\pa f}{\pa \tau} + \frac{p}{E}\hat{p}^i\frac{\pa f}{\pa x^i}+p\frac{\pa f}{\pa p}\left[-\mathcal{H}+\frac{\pa \phi}{\pa \tau}-\frac{E}{p}\hat{p}^i\frac{\pa \psi}{\pa x^i}   \right],
\ee
where in this work $\mathcal{H} = d\ln{a}/d\tau$ is the conformal Hubble expansion rate. For massless particles, it is generically simpler to introduce the comoving momentum $q\equiv a p$ and comoving energy $\epsilon \equiv a E$. In this case, the left-hand side of the Boltzmann equation can be written
\be
\frac{df}{d\tau} =\frac{\pa f}{\pa \tau} + \frac{q}{\epsilon}\hat{q}^i\frac{\pa f}{\pa x^i}+q\frac{\pa f}{\pa q}\left[\frac{\pa \phi}{\pa \tau}-\frac{\epsilon}{q}\hat{q}^i\frac{\pa \psi}{\pa x^i}   \right].
\ee
 Using Eq.~\eqref{eq:lambda}, the Boltzmann equations for DM and DR then take the form
 \be\label{eq:Gen_Boltz_DM}
 \frac{\pa \fD}{\pa \tau} + \frac{p}{E}\hat{p}^i\frac{\pa \fD}{\pa x^i}+p\frac{\pa \fD}{\pa p}\left[-\mathcal{H}+\frac{\pa \phi}{\pa \tau}-\frac{E}{p}\hat{p}^i\frac{\pa \psi}{\pa x^i}   \right]    = \frac{a}{E}(1+\psi)C_{\chi\tilde{\gamma}\leftrightarrow\chi\tilde{\gamma}}\left[\pp{}\right],
 \ee
 \be\label{eq:Gen_Boltz_DR}
 \frac{\pa \fR}{\pa \tau} + \frac{q}{\epsilon}\hat{q}^i\frac{\pa \fR}{\pa x^i}+q\frac{\pa \fR}{\pa q}\left[\frac{\pa \phi}{\pa \tau}-\frac{\epsilon}{q}\hat{q}^i\frac{\pa \psi}{\pa x^i}   \right] =\frac{a^2}{\epsilon}(1+\psi)\left(C_{\chi\tilde{\gamma}\leftrightarrow\chi\tilde{\gamma}}\left[\frac{\qq}{a}\right]+C_{\tilde{\gamma}\tilde{\gamma}\leftrightarrow\tilde{\gamma}\tilde{\gamma}}\left[\frac{\qq}{a}\right]\right).
 \ee
 We note that the only assumptions that went into deriving these equations is the perturbativity of the scalar gravitational potentials and that the mean distribution function is isotropic. In the following subsections, we further simplify these equations by assuming that the phase space distribution functions of DM and DR are nearly spatially homogenous and isotropic.
 %%%%
 \subsubsection{Dark Radiation}
 %%%%
 We assume that the distribution function of DR is close to its thermal equilibrium value and we parametrize the deviation from perfect equilibrium as follows
\be
\fR(\xx,\q,\tau)=\fR^{(0)}(q,\tau)[1+\Theta_{\rm DR}(\xx,\q,\tau)],
\ee
where $\fR^{(0)}(q,\tau)$ denotes the isotropic and homogeneous equilibrium DR distribution function which would be a Fermi-Dirac (Bose-Einstein) distribution for fermionic (bosonic) DR. Keeping only the terms that do not contain perturbed quantities in Eq.~\eqref{eq:Gen_Boltz_DR}, we obtain the zeroth-order Boltzmann equation for DR
\be
\frac{\pa \fR^{(0)}(q)}{\pa \tau} = \frac{a^2}{\epsilon} \left( C^{(0)}_{\chi\tilde{\gamma}\leftrightarrow\chi\tilde{\gamma}}[\fR^{(0)},\fD^{(0)}]+C^{(0)}_{\tilde{\gamma}\tilde{\gamma}\leftrightarrow\tilde{\gamma}\tilde{\gamma}}[\fR^{(0)}]\right),
\ee
where $\fD^{(0)}$ and $C^{(0)}$ denote the unperturbed (isotropic and homogeneous) DM distribution function and collision term, respectively. This equation essentially controls the kinetic energy transfer between the DM and the DR which, as long as it is efficient, will result in setting $T_\chi = T_{\rm DR}$ (more details in the dark matter subsection below). The first-order DR Boltzmann equation is 
\be
\fR^{(0)}\left[\frac{\pa \Theta_{\rm DR}}{\pa \tau}+i\frac{q}{\epsilon}k\mu\Theta_{\rm DR}\right]+q\frac{\pa\fR^{(0)}}{\pa q}\left[\frac{\pa \phi}{\pa\tau}-i\frac{\epsilon}{q}k\mu \psi\right]+ \frac{a^2C^{(0)}[q/a]}{\epsilon}(\Theta_{\rm DR}-\psi)=\frac{a^2}{\epsilon}C^{(1)}\left[\frac{\qq}{a}\right],
\ee
where we have taken a Fourier transform with respect to the wave number $k$, and where we have used the zeroth-order equation to simplify the above. It is understood that the perturbation variables are now evaluated in Fourier space and $\mu \equiv \hat{q}\cdot\hat{k}$, $k=|\kk|$, and $\hat{k}=\kk/k$. $C^{(1)}$ stands for the first-order collision term. For notational convenience  we have suppressed the sum over the different scattering channels; it is understood that $C^{(0)}$ and $C^{(1)}$ are summed over the different processes. Since we are focusing purely on (helicity) scalar fluctuations, we can expand the $\mu$-dependence of $\Theta$ in Legendre polynomials as follows
\be\label{eq:Legendre_expansion}
\Theta_{\rm DR}(k,\hat{q}, q,\tau)=\sum_{l=0}^{\infty}(-i)^l(2l+1)F_l(k,q,\tau)P_l(\mu).
\ee
Substituting the above expansion in the first-order Boltzmann equation and integrating both sides with $\frac{1}{2(-i)^l}\int_{-1}^1d\mu P_l(\mu)$ yields
\ba\label{eq:DR_massive}
\fR^{(0)}\left[\frac{\pa F_l}{\pa \tau}+k\frac{q}{\epsilon}\left(\frac{l+1}{2l+1}F_{l+1}-\frac{l}{2l+1}F_{l-1}\right)\right]+q\frac{\pa \fR^{(0)}}{\pa q}\left[\frac{\pa \phi}{\pa\tau}\de_{l0}+\frac{k}{3}\frac{\epsilon}{q}\psi\de_{l1}\right]&+&\\
\frac{a^2C^{(0)}[q/a]}{\epsilon}(F_l-\psi\de_{l0})&=&\frac{a^2}{\epsilon}\frac{1}{2(-i)^l}\int_{-1}^1d\mu P_l(\mu)C^{(1)}\left[\frac{\qq}{a}\right],\nonumber
\ea
where $\de_{ij}$ is the Kronecker delta. The above equation represents an infinite hierarchy of equations for the different multipole moments of the DR distribution. Omitting the collision terms, these equations are essentially those describing the cosmological evolution of, e.g., massive neutrinos, and have been extensively studied in the literature (see e.g. \cite{Ma:1995ey}). The addition of the collision integrals can lead to frequent scattering between DM and DR (or DR self-interaction) that prohibits DR free-streaming, hence suppressing all multipole moments with $l\geq2$ and leaving only the monopole ($l=0$) and dipole ($l=1$) to solve for. However, in models where DR eventually decouples from DM (or itself) the higher multipole moments become important and must be included in the computation.

The hierarchy of equations given in Eq.~\eqref{eq:DR_massive} is very general and can be used to describe the evolution of a large variety of interacting massive and massless DR models. In the present work, we exclusively focus on massless DR  ($\epsilon = q$) since it allows a dramatic simplification to the above equations. We emphasize that the ETHOS framework does not depend on this specific choice, and the formalism could easily be expanded to handle massive DR. We also neglect the term proportional to the zeroth-order collision term in Eq.~\eqref{eq:DR_massive}. This is usually a very good approximation since this term can only contribute when the DM-DR system significantly departs from thermal equilibrium\footnote{There are some instances where this term could play a role, such as in models where DM never fully reaches thermal equilibrium with DR, or in models where DM decays to DR.}. With these simplifications, we can rewrite Eq.~\eqref{eq:DR_massive} as

\be\label{eq:DR_massless_temp_pert}
\frac{\pa \nu_l}{\pa \tau}+k\left(\frac{l+1}{2l+1}\nu_{l+1}-\frac{l}{2l+1}\nu_{l-1}\right)-4\left[\frac{\pa \phi}{\pa\tau}\de_{l0}+\frac{k}{3}\psi\de_{l1}\right] = -\frac{a^2}{q}\frac{2}{(-i)^l}\frac{1}{\frac{\pa\fR^{(0)}}{\pa\ln q}}\int_{-1}^1d\mu P_l(\mu)C^{(1)}\left[\frac{\qq}{a}\right], 
\ee
where
\be\label{def:temp_perturbation}
\nu_l\equiv -4\frac{F_l}{\frac{\pa\ln\fR^{(0)}}{\pa\ln q}},
\ee
The variable $\nu_l$ is usually referred to as a temperature fluctuation since it corresponds to a local redefinition of the DR temperature. As we discuss below, expressing the DR hierarchy in terms of the $\nu_l$ variables also simplifies the structure of the collision term.

%%%% 
\subsubsection{Nonrelativistic Dark Matter}
%%%%
We shall now deviate from the complete generality of Eq.~\eqref{eq:Gen_Boltz_DM} and assume that DM is a stable particle that is nonrelativistic at all epochs of interest for structure formation. This implies that the term involving $p/E\sim p/m_\chi\ll1$ in Eq.~(\ref{eq:LHS_Boltzmann_eq}) can be considered a small perturbation. Neglecting these small perturbations, the zeroth order Boltzmann equation for DM can be written as
\be\label{eq:zerothDM}
\frac{\pa \fD^{(0)}}{\pa\tau}-p \mathcal{H} \frac{\pa\fD^{(0)}}{\pa p} = \frac{a}{E}C^{(0)}_{\chi\tilde{\gamma}\leftrightarrow\chi\tilde{\gamma}}[\fR^{(0)},\fD^{(0)}].
\ee
In analogy with the thermal case for nonrelativistic particles, we define the DM temperature as \cite{Bertschinger:2006nq,2007JCAP...04..016B}
\be\label{eq:T_and_number_density}
T_\chi \equiv \frac{\eta_\chi}{3 n_\chi^{(0)} } \int\frac{d^3p}{(2\pi)^3}\frac{\pp{}^2}{m_\chi}\fD^{(0)}(p),\qquad\text{where}\qquad n_\chi^{(0)} \equiv \eta_\chi\int\frac{d^3p}{(2\pi)^3}f_\chi^{(0)}(p),
\ee
where $m_\chi$ is the DM mass, $n_\chi^{(0)}$ is the homogeneous and isotropic DM number density, and where $\eta_\chi$ is the number of internal degrees of freedom of DM particles. We can multiply Eq.~(\ref{eq:zerothDM}) by $ \frac{\eta_\chi}{3 n_\chi^{(0)} }\int \frac{d^3p}{(2\pi)^3}\frac{\pp{}^2}{m_\chi}$, and integrate over $\pp{}$ to obtain the evolution equation of the DM temperature \citep{Bertschinger:2006nq}
\be\label{eq:DM_temperature_evol_gen}
\frac{dT_\chi}{d\tau} +2 \mathcal{H} T_\chi - \frac{a\,\eta_\chi}{3 n_\chi^{(0)} m_\chi}\int \frac{d^3p}{(2\pi)^3}\frac{\pp{}^2}{m_\chi}C^{(0)}_{\chi\tilde{\gamma}\leftrightarrow\chi\tilde{\gamma}} =0.
\ee
The second term on the left-hand side accounts for the adiabatic cooling of the DM due to the expansion of the Universe, while the third term accounts for the DR heating. As long as the heating rate is much larger than the Hubble expansion rate, the DM will be in thermal equilibrium with the DR and $T_\chi = T_{\rm DR}$.

We now turn our attention to the DM perturbations. For nonrelativistic DM, the exact form of the zeroth-order distribution function is almost exactly Maxwellian until just before 
kinetic decoupling \citep{2007JCAP...04..016B}. Just like for the zeroth order \cite{Bertschinger:2006nq,2007JCAP...04..016B}, the strategy to obtain the equation for the density and velocity fluctuations of DM is to take moments of Eq.~(\ref{eq:Gen_Boltz_DM}), keeping only the leading order terms in the small quantity $p/E\ll1$. We take the first moment of Eq.~(\ref{eq:Gen_Boltz_DM}):
\be\label{eq:DM_monopole_first_eq}
\int \frac{d^3p}{(2\pi)^3}\left[ \frac{\pa \fD}{\pa \tau} + \frac{p}{E}\hat{p}^i\frac{\pa \fD}{\pa x^i}+p\frac{\pa \fD}{\pa p}\left[-\mathcal{H}+\frac{\pa \phi}{\pa \tau}-\frac{E}{p}\hat{p}^i\frac{\pa \psi}{\pa x^i}   \right]\right]=\int \frac{d^3p}{(2\pi)^3}\left[\frac{a}{E}(1+\psi)C_{\chi\tilde{\gamma}\leftrightarrow\chi\tilde{\gamma}}\left[\pp{}\right]\right].
\ee
Recalling the definition of DM bulk velocity and total number density\footnote{We emphasize that the DM number density defined in Eq.~\eqref{eq:T_and_number_density} is different from that defined in Eq.~\eqref{eq:DM_bulk_velocity}; the former  is homogeneous across space while the latter depends on spatial position.},
\be\label{eq:DM_bulk_velocity}
 \vec{v}_\chi\equiv \frac{\eta_\chi}{n_\chi}\int\frac{d^3p}{(2\pi)^3}f_\chi(\pp{})\frac{p\,\hat{\bf p}}{E},\qquad n_\chi \equiv \eta_\chi\int\frac{d^3p}{(2\pi)^3}f_\chi (\pp{}),
\ee
we obtain
\be
\frac{\pa n_\chi}{\pa \tau}+3\mathcal{H} n_\chi+\frac{\pa (n_\chi v^i)}{\pa x^i}-3\frac{\pa \phi}{\pa\tau}n_\chi=0.
\ee
We note that the collision term has to be zero here since scattering alone cannot change the number density of dark matter (see Appendix \ref{sec_app:chichi_int} for details). Expanding the number density of dark matter as
\be
n_\chi(\xx,\tau)  \equiv n_\chi^{(0)}(\tau)[1+\de_\chi(\xx,\tau)],
\ee
where $n_\chi^{(0)}$ is defined in Eq.~\eqref{eq:T_and_number_density}, and where the above is used to define $\de_\chi$. Keeping only the first order pieces and performing a Fourier transform yields the equation
\be\label{eq:DM_density}
\dot{\de}_\chi+\theta_\chi-3\dot{\phi}=0,
\ee
where an overhead dot denotes a derivative with respect to conformal time, $\theta_\chi \equiv i \vec{k}\cdot \vec{v}_\chi$ is the divergence of the DM velocity, and where it is understood that the perturbation variables are evaluated in Fourier space. To close the dark matter system of equations, we need an equation for its bulk velocity. We multiply both sides of Eq.~(\ref{eq:Gen_Boltz_DM}) by $\frac{p\hat{\bf p}}{E}$ and integrate over all $\pp{}$
\be\label{eq:pre_vel}
\int \frac{d^3p}{(2\pi)^3}\frac{p\hat{\bf p}}{E}\left[ \frac{\pa \fD}{\pa \tau} + \frac{p}{E}\hat{p}^i\frac{\pa \fD}{\pa x^i}+p\frac{\pa \fD}{\pa p}\left[-\mathcal{H}+\frac{\pa \phi}{\pa \tau}-\frac{E}{p}\hat{p}^i\frac{\pa \psi}{\pa x^i}   \right]\right]=\int \frac{d^3p}{(2\pi)^3}\frac{p\hat{\bf p}}{E}\left[\frac{a}{E}(1+\psi)C_{\chi\tilde{\gamma}\leftrightarrow\chi\tilde{\gamma}}\left[\pp{}\right]\right].
\ee
Let us first compute the left-hand side of the equation. To first order in perturbation theory, we obtain
\be
\frac{\pa (n_\chi^{(0)} \vec{v}_\chi)}{\pa\tau}+\vec{\nabla}\left[\frac{\de p_\chi}{m_\chi}\right]+4 \mathcal{H} n_\chi^{(0)}  \vec{v}_\chi +n_\chi^{(0)}  \vec{\nabla}\psi,
\ee
where $\de p_\chi$ stands for the dark matter pressure perturbation. Assuming that the pressure perturbation is adiabatic, we can write $\de p_\chi = c_{\chi}^2 \de \rho_\chi = c_{\chi}^2 n_\chi^{(0)} m_\chi \de_\chi$, where $c_{\chi}^2$ is the dark matter sound speed squared. Using the fact that $\pa(a^3n_\chi^{(0)})/\pa\tau = 0$, the left-hand side of Eq.~(\ref{eq:pre_vel}) becomes
\be
\frac{\pa (\vec{\nabla}\cdot\vec{v}_\chi)}{\pa\tau}+c_{\chi}^2\nabla^2\de_\chi+\mathcal{H} \vec{\nabla}\cdot\vec{v}_\chi + \nabla^2\psi,
\ee
where we have multiplied the equation by the Nabla operator since only the divergence of the velocity couples to the scalar gravitational potential. The dark matter sound speed can be computed via
\be\label{eq:DM_sound_speed}
c_{\chi}^2 = \frac{\dot{P}_\chi}{\dot{\rho}_\chi}=\frac{\dot{n}^{(0)}_\chi T_\chi +n^{(0)}_\chi \dot{T}_\chi}{m_\chi \dot{n}^{(0)}_\chi }=\frac{3 \mathcal{H} T_\chi -\dot{T}_\chi}{3 \mathcal{H} m_\chi  } = \frac{T_\chi}{m_\chi}\left(1-\frac{\dot{T}_\chi}{3 \mathcal{H}  T_\chi}\right).
\ee
Finally, in Fourier space, the equation for the dark matter bulk velocity is
\be\label{eq:DM_velocity}
\dot{\theta}_\chi-c_{\chi}^2k^2\de_\chi+\mathcal{H} \theta_\chi - k^2\psi =\frac{a(1+\psi)\eta_\chi}{n_\chi^{(0)}} \int \frac{d^3p}{(2\pi)^3}\frac{p(i\vec{k}\cdot\hat{\bf p})}{E^2}C_{\chi\tilde{\gamma}\leftrightarrow\chi\tilde{\gamma}}\left[\pp{}\right].
\ee
Together with the Einstein equations, Eqs.~\eqref{eq:DR_massless_temp_pert}, \eqref{eq:DM_density}, and \eqref{eq:DM_velocity} fully describe the scalar cosmological fluctuations of nonrelativistic DM coupled to a relativistic component. We emphasize that these equations are general and can be used to determine the evolution of dark matter fluctuations for a broad range of particle models, whose specific details only enter the problem via the collision integrals. As such, they form the backbone of the ETHOS framework as applied to nonrelativistic DM interacting with some form of relativistic DR.  

%%%%%
\subsection{Structure of the DM-DR collision term}\label{sec:coll_int}
%%%%%

We now turn our attention to the general structure of the collision integrals appearing in Eqs~\eqref{eq:DR_massless_temp_pert}, \eqref{eq:DM_temperature_evol_gen}, and \eqref{eq:DM_velocity}  for the elastic scattering process $\tilde{\gamma}(\pp{1})+\chi(\pp{2})\leftrightarrow \tilde{\gamma}(\pp{3})+\chi(\pp{4})$ \citep{Kolb:1990vq}

\be\label{eq:coll_Boltz_master}
C_{\chi\tilde{\gamma}\leftrightarrow\chi\tilde{\gamma}}[\pp{1}]= \frac{1}{2}\eta_\chi \int d\Pi_2 d\Pi_3 d\Pi_4 \left(\frac{1}{\eta_\chi\eta_{\rm DR}}\sum_{\rm states}|\mathcal{M}|^2\right)(2\pi)^4\de^4(P_1+P_2-P_3-P_4)F(\pp{1},\pp{2},\pp{3},\pp{4})
\ee
where $|\mathcal{M}|^2$ is the square of the matrix element for the scattering, $\eta_{\rm DR}$ is the DR degeneracy factor, $P_i$ denotes the $i^{\rm th}$ four-momentum, $\pp{i}$ denotes the $i^{\rm th}$ three-momentum, $p_i = |\pp{i}|$, and where
\be\label{eq:integral_measure}
 d\Pi_i=\frac{d^3p_i}{(2\pi)^32E_i},
 \ee
is the Lorentz invariant phase-space measure, and
\be
F(\pp{1},\pp{2},\pp{3},\pp{4})=\fc(\pp{4})\fR(\pp{3})(1\pm\fc(\pp{2}))(1\pm\fR(\pp{1}))-\fc(\pp{2})\fR(\pp{1})(1\pm\fc(\pp{4}))(1\pm\fR(\pp{3})),
\ee
where the ``$+$'' signs are for bosonic species and ``$-$'' signs are for fermionic species. We note that the leading factor of $1/2$ in Eq.~\eqref{eq:coll_Boltz_master} is necessary in order to ensure that both sides of the Boltzmann equation transform consistently under a Lorentz transformation. We emphasize that this factor is unrelated to possible symmetries between identical particles in the initial and final states. The details of the particle model enter through the matrix element squared 
\be
\frac{1}{\eta_\chi\eta_{\rm DR}}\sum_{\rm states}|\mathcal{M}|^2,
\ee
where the sum runs over all internal degrees of freedom (spins, colors, etc.). The prefactor is responsible for averaging over the initial states. Finally, the leading factor of $\eta_\chi$ in Eq.~\eqref{eq:coll_Boltz_master} is necessary since we need to sum over all the DM states that a DR particle can scatter on. Since we are only interested in situations with highly nonrelativistic DM, one can neglect the Pauli blocking/Bose enhancement factors for the $\chi$ particles
\be
F(\pp{1},\pp{2},\pp{3},\pp{4})\approx\fc(\pp{4})\fR(\pp{3})(1\pm\fR(\pp{1}))-\fc(\pp{2})\fR(\pp{1})(1\pm\fR(\pp{3})).
\ee
At early times, frequent scattering between DR and DM keeps these constituents in thermal equilibrium at a common temperature $T_{\rm DR}$, and the DR phase-space distribution assumes a Fermi-Dirac or Bose-Einstein shape, depending on whether DR is a fermion or a boson. Similar to the case of neutrinos decoupling from the Standard Model plasma \citep{Kolb:1990vq}, the DR phase-space distribution stays very close to its equilibrium configuration after DR decoupling, with its temperature scaling as $T_{\rm DR}\propto 1/a$. Residual thermal coupling between DM and DR\footnote{This is similar to the case of CMB photons, which remain thermally coupled to baryons until long after the epoch of CMB last scattering.} can however imprint small spectral distortions on the DR spectrum. Indeed, DM effectively acts as a heat sink for DR since the latter must constantly provide energy to the DM particles such that their temperature tracks that of the DR ($T_\chi=T_{\rm DR}\propto 1/a$) instead of cooling adiabatically like $T_\chi\propto1/a^2$. For ultrarelativistic DR scattering off nonrelativistic DM, we naturally expect the integrated magnitude of these spectral distortions to be of the order of the DM to DR entropy ratio. Up to a factor of order unity, the overall DR energy loss to DM heating is 
\be
\frac{\Delta\rho_{\tilde{\gamma}}}{\rho_{\tilde{\gamma}}}\sim 10^{-8}\left(\frac{\Omega_\chi h^2}{0.12}\right)\left(\frac{m_\chi}{\text{GeV}}\right)^{-1}\left(\frac{\xi}{0.5}\right)^{-3},
\ee
where $\rho_{\tilde{\gamma}}$ is the DR energy density, $\Omega_\chi$ is the DM density in units of the critical density of the Universe, and $\xi \equiv (T_{\rm DR}/T_{\rm CMB})|_{z=0}$ is the present-day DR to CMB temperature ratio (assuming $T_\mathrm{DR}\propto 1/a$ until today). Thus, as long as the DR is not abnormally cold compared to the visible sector, spectral distortions will remain small and it is an excellent approximation to take the zeroth-order DR phase-space distribution function to be either Fermi-Dirac or Bose-Einstein
\be\label{eq:background_DR_dist_func}
\fR^{(0)}(\pp{}) = \frac{1}{e^{(p/T_{\rm DR})}\pm1},
\ee
where we have assumed no chemical potential. For massless DR interacting with nonrelativistic DM, it is important to realize that little momentum is exchanged in a typical DM-DR scattering process. This is similar to the familiar case of Thomson scattering of CMB photons off free electrons near the epoch of cosmological recombination. Quantitatively, the typical momentum of a DR particle is $p_{\rm DR}\sim T_{\rm DR}$, while that of a DM particle is  $p_\chi \sim \sqrt{m_\chi T_\chi}$, immediately implying $p_{\rm DR}\ll p_\chi$ for nonrelativistic DM with $T_\chi \sim T_{\rm DR}$. Therefore, the change to the momentum of a DM particle from a single DR collision is very small. As we discuss below, this allows us to simplify the computation of the collision term by doing a systematic expansion in the small momentum transfer (see e.g. \cite{2007JCAP...04..016B}). 
\subsubsection{Zeroth-order collision term}
The zeroth order collision term was computed in detail in Ref.~\cite{2007JCAP...04..016B} (see also~\cite{Bertschinger:2006nq}), and we thus simply quote their results here:
\ba\label{Eq:zeroth_order_coll_term_DM}
C^{(0)}_{\chi\tilde{\gamma}\leftrightarrow\chi\tilde{\gamma}}[\pp{}] &=&\frac{\eta_{\rm DR}}{12(2\pi)^3m_\chi^2}\int dq\, \fR^{(0)}(q) \frac{\pa}{\pa q} \left[ q^4 \left(\frac{1}{\eta_\chi\eta_{\rm DR}}\sum_{\rm states} |\mathcal{M}|^2\right)\Bigg{|}_{\begin{subarray}{l} t=0 \\ s=m_\chi^2+2q m_\chi \end{subarray}}\right]\left(m_\chi T_{\rm DR} \nabla_{\pp{}}^2+\pp{}\cdot\nabla_{\pp{}}+3\right)\fD^{(0)}(p)\en
&=& \frac{\eta_{\rm DR}}{12(2\pi)^3} m_\chi^2 \left[\sum_n c_n (n+4)!\zeta(n+4)\gamma_n \left(\frac{T_{\rm DR}}{m_\chi}\right)^{n+4}\right]\left(m_\chi T_{\rm DR} \nabla_{\pp{}}^2+\pp{}\cdot\nabla_{\pp{}}+3\right)\fD^{(0)}(p)\label{eq:zeroth_order_coll_DM_Temp},
\ea
where we have expanded the matrix element as
\be\label{eq:momentum_power_law}
\left(\frac{1}{\eta_\chi\eta_{\rm DR}}\sum_{\rm states} |\mathcal{M}|^2\right)\Bigg{|}_{\begin{subarray}{l} t=0 \\ s=m_\chi^2+2q m_\chi \end{subarray}} = \sum_n c_n\left(\frac{q}{m_\chi}\right)^n,
\ee
and where $\gamma_n = (1-2^{-n-3})$ for fermionic DR and $\gamma_n=1$ for bosonic DR. Here $\zeta(z)$ is the Riemann Zeta function. It is important to notice that the incoming momentum $\pp{}$ in Eq.~\eqref{Eq:zeroth_order_coll_term_DM} denotes the DM momentum, while the incoming $\pp{1}$ in Eq.~\eqref{eq:coll_Boltz_master} stands for the DR momentum. We note that the above result was achieved by performing an expansion in the small momentum transfer exchanged in a typical DR-DM collision. Its generalization to scattering with DR particles that are not ultrarelativistic is tedious but straightforward \citep{Bringmann:2009vf}. Note that the same expression holds even if the amplitude is not Taylor expandable around vanishing momentum transfer $t=0$, but $|\mathcal{M}|^2$ should then be {\it averaged} over $t$ rather than evaluated at $t=0$ \citep{Kasahara,Gondolo:2012vh}.
\subsubsection{First-order collision term}\label{app:sec:First-order collision term}
We now turn our attention to the part of the collision integrals that is first order in the small perturbation variables $\nu_l$ and $v_\chi$. The computation is somewhat similar to that usually performed for CMB photons scattering off electrons, but it is more general since we allow for more complex momentum and angular dependence of the DM-DR scattering cross section.  Keeping only the first order\footnote{We note that this expression seems to explicitly contain zeroth order terms, but these exactly cancel out and do not contribute to $C^{(1)}_{\chi\tilde{\gamma}\leftrightarrow\chi\tilde{\gamma}}$.} terms in the perturbation variable $\Theta_{\rm DR}$, we can rewrite the collision term given in Eq.~\eqref{eq:coll_Boltz_master} as:
\ba
C^{(1)}_{\chi\tilde{\gamma}\leftrightarrow\chi\tilde{\gamma}}[\pp{1}]&=&\frac{1}{2}\eta_\chi \int d\Pi_2 d\Pi_3 d\Pi_4\left(\frac{1}{\eta_\chi\eta_{\rm DR}}\sum_{\rm states} |\mathcal{M}|^2\right)(2\pi)^4\de^4(P_1+P_2-P_3-P_4)\fR^{(0)}(\pp{1})\fR^{(0)}(\pp{3})\\
&&\qquad\times\left(\fc(\pp{4})\left(e^{p_1/T_{\rm DR}}(\Theta_{\rm DR}(\pp{3})+1)\pm\Theta_{\rm DR}(\pp{1})\right)-\fc(\pp{2})\left(e^{p_3/T_{\rm DR}}(\Theta_{\rm DR}(\pp{1})+1)\pm\Theta_{\rm DR}(\pp{3})\right)\right).\nonumber
\ea
We use the space part of the delta function to perform the $\pp{4}$ integral. The DM is assumed to be highly nonrelativistic and we can thus write $E_\chi \approx m_\chi + p_\chi^2/(2m_\chi)$. We use the fact that little momentum is exchanged in a typical collision to expand the delta function as
\be\label{delta_expansion}
\de\left(p_1+\frac{p_2^2}{2m_\chi}-p_3-\frac{(\pp{1}+\pp{2}-\pp{3})^2}{2m_\chi}\right)\simeq \de(p_1-p_3)+ \frac{(\pp{1}-\pp{3})\cdot \pp{2}}{m_\chi}\frac{\pa \de(p_1-p_3)}{\pa p_3},
\ee
where the derivative of the Dirac delta function is defined via integration by parts. The first term in Eq.~(\ref{delta_expansion}) yields
\ba\label{eq:zeroth_q_pre}
\frac{\pi}{ 2m_\chi}\fR^{(0)}(\pp{1})\eta_\chi \int d\Pi_2 d\Pi_3 \left(\frac{1}{\eta_\chi\eta_{\rm DR}}\sum_{\rm states} |\mathcal{M}|^2\right)\de(p_1-p_3)\fR^{(0)}(\pp{3})\hspace{8cm}\\
\times\left(\fc(\pp{1}+\pp{2}-\pp{3})\left(e^{p_1/T_{\rm DR}}(\Theta_{\rm DR}(\pp{3})+1)\pm\Theta_{\rm DR}(\pp{1})\right)-\fc(\pp{2})\left(e^{p_3/T_{\rm DR}}(\Theta_{\rm DR}(\pp{1})+1)\pm\Theta_{\rm DR}(\pp{3})\right)\right). \nonumber
\ea
Since $p_2\gg p_1,p_3$, it is a good approximation to write $\fc(\pp{1}+\pp{2}-\pp{3})\simeq \fc(\pp{2})$. We can now perform the $p_3$ integral and Eq.~(\ref{eq:zeroth_q_pre}) reduces to
\be\label{eq:coll1_with_mat_elem_ang_dep}
\frac{p_1}{8(2\pi)^2 m_\chi} \eta_\chi \fR^{(0)}(\pp{1})\int d\Pi_2 \fc(\pp{2})  \int d\Omega_3 \left(\frac{1}{\eta_\chi\eta_{\rm DR}}\sum_{\rm states} |\mathcal{M}|^2\right)\Bigg{|}_{t=2p_1^2(\tilde{\mu}-1)}\Big[\Theta_{\rm DR}(p_1\hat{\bf p}_3)-\Theta_{\rm DR}(\pp{1})\Big].
\ee
Here, we have computed the matrix element evaluated at momentum transfer $t=2p_1^2(\tilde{\mu}-1)$, where $\tilde{\mu} = \hat{\bf p}_1\cdot\hat{\bf p}_3$. To make further progress in evaluating the remaining integrals, we need to examine the structure of the matrix element. Writing the latter in terms of the Mandelstam variable $t$ and $s = m_\chi^2 +2 p_1 m_\chi(1-(p_2/m_\chi)\hat{\bf p}_1\cdot\hat{\bf p}_2)$, we note that the dependence on the incoming scattering angle of the cross section always appears multiplied by the quantity $p_2/m_\chi\ll1$ \citep{2007JCAP...04..016B}. Since the squared matrix element in Eq.~\eqref{eq:coll1_with_mat_elem_ang_dep} is multiplied by the small perturbations $\Theta_{\rm DR}$, we can neglect the dependence of the matrix element on the angle between the incoming particles since they would lead to second-order terms. A similar argument allows us to neglect the $p_2$ dependence of the matrix element. In order to perform the angular integration over $d\Omega_3$, we expand the $\tilde{\mu}$ dependence of the matrix element in Legendre polynomials,
\be\label{eq:exp_of_the_matrix_element_in_Leg}
\left(\frac{1}{\eta_\chi\eta_{\rm DR}}\sum_{\rm states} |\mathcal{M}|^2\right)\Bigg{|}_{\begin{subarray}{l} t=2p_1^2(\tilde{\mu}-1) \\ s=m_\chi^2+2p_1m_\chi \end{subarray}}=\sum_{n=0}^\infty (2n+1)A_n(p_1) P_n(\tilde{\mu}).
\ee
Equation \eqref{eq:coll1_with_mat_elem_ang_dep} then becomes
\be
 \frac{p_1}{16(2\pi)^2 m_\chi^2} n_\chi^{(0)} \fR^{(0)}(\pp{1}) \int d\Omega_3 \left[\sum_{n=0}^\infty (2n+1)A_n(p_1) P_n(\tilde{\mu})\right]\left[\sum_{l=0}^\infty (-i)^l(2l+1)F_l(p_1)\left(P_l(\hat{\bf p}_3\cdot\hat{\bf k})-P_l(\mu)\right)\right],
\ee
where we used Eq.~\eqref{eq:T_and_number_density} for the DM number density as well as Eq.~\eqref{eq:Legendre_expansion} to write down the angular dependence of the $\Theta_{\rm DR}$ variables. The azimuthal integration can be performed using the identity
\be\label{eq:Leg_P_identity}
\int_0^{2\pi} d\phi P_l(\hat{\bf p}_3\cdot\hat{\bf k}) = 2\pi P_l(\hat{\bf p}_1\cdot\hat{\bf p}_3)P_l(\hat{\bf p}_1\cdot\hat{\bf k})
\ee
where $\phi$ is the angle for $\hat{\bf p}_3$ to wrap around $\hat{\bf p}_1$. We can now use the orthogonality of the Legendre polynomials to perform the $\tilde{\mu}$ integral,
\be
\frac{p_1}{16\pi m_\chi^2} n_\chi^{(0)}  \fR^{(0)}(\pp{1})  \left[\sum_{l=1}^\infty (-i)^l(2l+1)F_l(p_1)P_l(\mu)\left(A_l(p_1)-A_0(p_1)\right)\right].
\ee
We note that the DR monopole ($l=0$) drops out of the problem since scattering alone cannot modify it. We now turn our attention to the second term of Eq.~(\ref{delta_expansion}). Since the quantity $p_2/m_\chi$ is itself a very small quantity, it is therefore sufficient to only keep terms that are zeroth order in the $\Theta_{\rm DR}$ variables
\be
\frac{\pi\eta_\chi}{2m_\chi}\fR^{(0)}(\pp{1})\int d\Pi_2 \fc(\pp{2}) \int d\Pi_3  \fR^{(0)}(\pp{3}) \left(\frac{1}{\eta_\chi\eta_{\rm DR}}\sum_{\rm states} |\mathcal{M}|^2\right)\frac{(\pp{1}-\pp{3})\cdot \pp{2}}{m_\chi}\frac{\pa \de(p_1-p_3)}{\pa p_3}\left(e^{p_1/T_{\rm DR}}-e^{p_3/T_{\rm DR}}\right).
\ee
As mentioned above, the matrix element is independent of $\pp{2}$ to leading order in the small quantity $p_2/m_\chi$. Carrying out the $\pp{2}$ integration yields the dark matter bulk velocity [see Eq.~\eqref{eq:DM_bulk_velocity} above]. We also use integration by parts to perform the $p_3$ integral, which yields
\be
-\frac{p_1}{16 (2\pi)^2m_\chi^2} n_\chi^{(0)} \frac{\pa \fR^{(0)}(\pp{1})}{\pa \ln p_1}  \int d\Omega_3 (\hat{\bf p}_1-\hat{\bf p}_3)\cdot\vec{v}_\chi \left(\frac{1}{\eta_\chi\eta_{\rm DR}}\sum_{\rm states} |\mathcal{M}|^2\right)\Bigg{|}_{\begin{subarray}{l} t=2p_1^2(\tilde{\mu}-1) \\ s=m_\chi^2+2p_1m_\chi \end{subarray}}.
\ee
Since we are focusing uniquely on scalar cosmological perturbations in the present work, we only need to consider the irrotational part of the DM velocity. This immediately implies that $\hat{\bf p}_1\cdot\vec{v}_\chi = \mu\, v_\chi =P_1(\mu)\, v_\chi $ and $\hat{\bf p}_3\cdot\vec{v}_\chi = P_1(\hat{\bf p}_3\cdot\hat{\bf k})v_\chi$. We can use the expansion of the matrix element given in Eq.~\eqref{eq:exp_of_the_matrix_element_in_Leg} together with the identity given in Eq.~\eqref{eq:Leg_P_identity} to perform the remaining $d\Omega_3$ integration,
\be
-\frac{p_1}{16 \pi m_\chi^2} n_\chi^{(0)} \frac{\pa \fR^{(0)}(\pp{1})}{\pa \ln p_1}P_1(\mu)\left(A_0(p_1) -A_1(p_1)\right)  v_\chi .
\ee
Using the definition of the DR temperature perturbation $\nu_l$ [see Eq.~\eqref{def:temp_perturbation}], the total first-order collision term for $\tilde{\gamma}(\pp{1})+\chi(\pp{2})\leftrightarrow \tilde{\gamma}(\pp{3})+\chi(\pp{4})$ scattering is thus
\be\label{final_coll_term}
C^{(1)}_{\chi\tilde{\gamma}\leftrightarrow\chi\tilde{\gamma}}[\pp{1}] =\frac{p_1}{16\pi m_\chi^2}  n_\chi^{(0)} \frac{\pa \fR^{(0)}(\pp{1})}{\pa\ln p_1} \Big[\frac{1}{4}\sum_{l=1}^{\infty}(-i)^l(2l+1)\nu_l(p_1)P_l(\mu)\left(A_0(p_1)-A_l(p_1)\right)-P_1(\mu)\left(A_0(p_1) - A_1(p_1)\right)  v_\chi \Big].
\ee
Here, the advantage of the $\nu_l$ variables becomes evident: they are the quantities that couple directly to the DM velocity without extra derivative of the background DR distribution function. 
%%%%
\subsection{Structure of the DR-DR collision term}\label{sec:DRDR_coll}
%%%%
In many models where DM interacts with some form of DR, it is possible for the latter to also have self-interactions via a process of the form $\tilde{\gamma}\tilde{\gamma}\leftrightarrow\tilde{\gamma}\tilde{\gamma}$. In general, the presence of DR self-interaction would only have a small impact on the evolution of the DM density field, but we nevertheless include it here since our goal is to develop a complete and self-consistent framework. Since the momentum exchanged in a typical DR-DR scattering event is of order unity, the computation of the collision integrals are significantly more complex than in the case of DM-DR scattering. We refer the reader to Ref.~\cite{Oldengott:2014qra} for a detailed exposure of the subtleties involved in accurately computing the self-interaction collision term for massless DR. Since we are mainly interested here in computing the DM power spectrum and not in the details of the DR spectrum, we adopt a simplified picture of DR-DR scattering in which we assume that the DR perturbation variables $\nu_l(p)$ are independent of the momentum $p$. This is equivalent to assuming that the DR spectrum remains purely thermal throughout the evolution of the Universe, and it is consistent with the choice made in Eq.~\eqref{eq:background_DR_dist_func}. We expand more on the validity of this assumption in Sec.~\ref{sec:DR_equations}. In this thermal approximation, the first-order DR-DR collision term admits the general form
\be\label{eq:DR-DR_collision_term}
C^{(1)}_{\tilde{\gamma}\tilde{\gamma}\leftrightarrow\tilde{\gamma}\tilde{\gamma}}[\pp{1}] =  p_1\frac{\pa \fR^{(0)}(\pp{1})}{\pa\ln p_1} \Lambda_{\tilde{\gamma}\tilde{\gamma}\leftrightarrow\tilde{\gamma}\tilde{\gamma}}(\pp{1}) \frac{1}{4}\sum_{l=1}^\infty (-i)^l (2l+1)P_l(\mu) \nu_l \left(1 - G_l(\pp{1})\right),
\ee
where the functions $ \Lambda_{\tilde{\gamma}\tilde{\gamma}\leftrightarrow\tilde{\gamma}\tilde{\gamma}}$ and $G_l$ encode the details of the DR self-interaction. We note that energy conservation implies that the $l=0$ mode exactly vanish in the above expansion. Similarly, momentum conservation within the DR fluid immediately implies that $G_1(\pp{1}) =1$.  Physically, the main effect of DR self-interaction is to suppress its free-streaming, which could in turn modify the diffusion (Silk) damping that DR imparts on the DM matter power spectrum.
%%%%
\subsection{Dark radiation equations}\label{sec:DR_equations}
%%%%
We can now substitute Eqs.~\eqref{final_coll_term} and \eqref{eq:DR-DR_collision_term} in Eq.~\eqref{eq:DR_massless_temp_pert} and use the orthogonality of the Legendre polynomials to perform the $\mu$ integral:
\ba\label{eq:nu_l_tot_eq_before_int}
\frac{\pa \nu_l}{\pa \tau}+k\left(\frac{l+1}{2l+1}\nu_{l+1}-\frac{l}{2l+1}\nu_{l-1}\right)-4\left[\frac{\pa \phi}{\pa\tau}\de_{l0}+\frac{k}{3}\psi\de_{l1}\right] =\hspace{8cm} \\
-a \frac{1}{16\pi m_\chi^2} n_\chi^{(0)} \left[\left(A_0(\frac{q}{a}) - A_l(\frac{q}{a})\right) \nu_l -\frac{4}{3}i v_\chi \left(A_0(\frac{q}{a})-A_1(\frac{q}{a})\right)\de_{l1} \right]-a\Lambda_{\tilde{\gamma}\tilde{\gamma}\leftrightarrow\tilde{\gamma}\tilde{\gamma}}(\frac{q}{a})\left(1-G_l(\frac{q}{a})\right)\nu_l .\nonumber 
\ea
As noted above, the right-hand side exactly vanishes for the monopole. In principle, one could solve this hierarchy of differential equations on a grid of $q$ values to obtain the complete solution $\nu_l(k,q,\tau)$, which can then be used to compute the physical quantities entering the perturbed Einstein equations. For massless DR,  the energy perturbation $\de_{\rm DR}$, the divergence of the DR velocity $\theta_{\rm DR}$, and the higher moments of the DR Boltzmann hierarchy $\Pi_l(k,\tau)$ are related to the $F_l(k,q,\tau)$ variables\footnote{We thank Manuel A. Buen-Abad for pointing out an inconsistency with these definitions in an earlier version of the manuscript.} as \citep{Ma:1995ey}
\be\label{pert_def}
\de_{\rm DR}(k,\tau) = \frac{\int dq\, q^3 \fR^{(0)}(q,\tau)F_0(k,q,\tau)}{\int dq\, q^3 \fR^{(0)}(q,\tau)},\quad \theta_{\rm DR}(k,\tau)= \frac{3}{4}k\frac{\int  dq \, q^3 \fR^{(0)}(q,\tau)F_1(k,q,\tau)}{\int dq\,q^3  \fR^{(0)}(q,\tau)},
\ee
\be
\Pi_{{\rm DR,}l}(k,\tau) = \frac{ \int dq\, q^3  \fR^{(0)}(q,\tau)F_l(k,q,\tau)}{\int dq\, q^3  \fR^{(0)}(q,\tau)},
\ee
respectively. We note that the DR shear perturbation is given by $\sigma_{\rm DR}(k,\tau)=\Pi_{{\rm DR,}2}(k,\tau)/2$. In practice however, it is much simpler to first integrate Eq.~\eqref{eq:nu_l_tot_eq_before_int} with respect to $q$ before solving the differential equations for the different $l$-moments. Indeed, the left-hand side of Eq.~\eqref{eq:nu_l_tot_eq_before_int} can straightforwardly be expressed in terms of the physical DR variables by multiplying it by $\int dq\,q^3 \fR^{(0)}(q)$, performing the $q$ integration, and dividing the result by $\int dq\,q^3 \fR^{(0)}(q)$. However, since the matrix element coefficients $A_l$ appearing on the right-hand side of Eq.~\eqref{eq:nu_l_tot_eq_before_int} depend on momentum, the collision term cannot in general be expressed directly in terms of the physical DR variables.\footnote{In the CMB case, the Thomson scattering matrix element is independent of momentum and the collision term can exactly be expressed in terms of physical variables.} In the present work, we assume that the DR spectrum remains exactly thermal throughout the evolution of the Universe, which immediately implies that the $\nu_l$ variables must be independent of $q$. For models where DM is in kinetic equilibrium with the DR at early times, this \emph{thermal} approximation is extremely good since the large scattering rate appearing in Eq.~\eqref{eq:nu_l_tot_eq_before_int} suppresses the $q$-dependence of the $\nu_l$ variables. For instance, frequent scattering events set $\nu_1(k,\tau) = (4/3)iv_\chi$ and $\nu_{l\geq2}(k,\tau)=0 $ at early times, independently of $q$. As the scattering rate becomes comparable to the Hubble expansion rate, the DR perturbation variables $\nu_l$ can develop a small $q$-dependence of the order of the DM to DR entropy ratio. In the following, we neglect this small correction since it has a negligible impact on the DM distribution at late times. Applying to the right-hand side the same operations that we performed on the left-hand side of Eq~\eqref{eq:nu_l_tot_eq_before_int} leads to the following hierarchy of equations: 
\ba
\dot{\de}_{\rm DR} +\frac{4}{3}\theta_{\rm DR}-4\dot{\phi}&=&0,\label{eq:DR_continuity}\\
\dot{\theta}_{\rm DR}+k^2(\sigma_{\rm DR}-\frac{1}{4}\de_{\rm DR})-k^2\psi&=&\dot{\kappa}_{\rm DR-DM} \,(\theta_{\rm DR}-\theta_{\chi}),\label{eq:DR_Euler}\\
\dot{\Pi}_{{\rm DR},l} + \frac{k}{2l+1}\left((l+1)\Pi_{{\rm DR},l+1} - l\Pi_{{\rm DR},l-1}\right)&=&\left(\alpha_l\dot{\kappa}_{\rm DR-DM} + \beta_l \dot{\kappa}_{\rm DR-DR}\right)\,\Pi_{{\rm DR},l},\label{eq:DR_higher_l_moments}
\ea
where $\dot{\kappa}_{\rm DR-DM}$ is the DR opacity to DM scattering,
\begin{align}\label{eq:dark_rad_opacity_def}
\dot{\kappa}_{\rm DR-DM}& \equiv \frac{a}{16\pi m_\chi^2} n_\chi^{(0)} \frac{\int dp\, p^4 \frac{\pa\fR^{(0)}(p)}{\pa p} \left[ A_0(p) - A_1(p)\right]}{\int dp\, p^3 \fR^{(0)}(p)},\en
&=\left(\frac{3n_\chi^{(0)}m_\chi}{4\rho_{\rm DR}}\right)\frac{a}{16\pi m_\chi^3}   \frac{\eta_{\rm DR}}{3}\int \frac{p^2dp}{2\pi^2}\, p^2 \frac{\pa\fR^{(0)}(p)}{\pa p}\left[ A_0(p) - A_1(p)\right] ,
\end{align}
where $\rho_{\rm DR} = \eta_{\rm DR} \zeta \pi^2 T_{\rm DR}^4/30$ with $\zeta=1$ for bosonic DR and $\zeta=7/8$ for fermionic DR, and where $\alpha_l$ are $l$-dependent coefficients that encompass information about the angular dependence of the DM-DR scattering cross section. They are given by
\be\label{eq:def_ang_coefficients_DR}
\alpha_l \equiv \frac{\int dp\, p^4 \frac{\pa\fR^{(0)}(p)}{\pa p}\left[A_0(p) -A_l(p)\right]}{\int dp\, p^4 \frac{\pa\fR^{(0)}(p)}{\pa p}\left[A_0(p) -A_1(p)\right]}.
\ee
Similarly, $\dot{\kappa}_{\rm DR-DR}$ is the DR opacity to self-scattering, which we write as
\be\label{eq:dark_rad_self_opacity_def}
\dot{\kappa}_{\rm DR-DR}=\frac{a\,\eta_{\rm DR}}{4\rho_{\rm DR}}\int \frac{p^2dp}{2\pi^2} p^2 \frac{\pa\fR^{(0)}(p) }{\pa p}\Lambda_{\tilde{\gamma}\tilde{\gamma}\leftrightarrow\tilde{\gamma}\tilde{\gamma}}(p),
\ee
and where we define the angular coefficients for DR-DR scattering as
\be\label{eq:def_ang_coefficients_DR-DR}
\beta_ l \equiv \frac{\int dp\, p^4 \frac{\pa\fR^{(0)}(p)}{\pa p} \Lambda_{\tilde{\gamma}\tilde{\gamma}\leftrightarrow\tilde{\gamma}\tilde{\gamma}}(p)\left(1-G_l(p)\right)}{\int dp\, p^4 \frac{\pa\fR^{(0)}(p)}{\pa p} \Lambda_{\tilde{\gamma}\tilde{\gamma}\leftrightarrow\tilde{\gamma}\tilde{\gamma}}(p)}.
\ee
%

%%%
\subsection{Dark matter equations}\label{sec:DM_eqs}
%%%
\subsubsection{Temperature and sound speed evolution}
%%%
Substituting the zeroth order collision term given in Eq.~\eqref{eq:zeroth_order_coll_DM_Temp}  into the evolution equation for the DM temperature [Eq.~(\ref{eq:DM_temperature_evol_gen})] 
\be\label{eq:DM_temperature_evol_final}
\frac{dT_\chi}{d\tau} +2 \mathcal{H} T_\chi - \Gamma_{\rm heat}(T_{\rm DR})\left(T_{\rm DR} - T_\chi\right)=0\,,
\ee
where the heating rate is

\be\label{eq:Gamma_heating}
\Gamma_{\rm heat}(T_{\rm DR}) = a \frac{\eta_{\rm DR}m_\chi}{6 (2\pi)^3}  \left[\sum_n c_n (n+4)!\zeta(n+4)\gamma_n \left(\frac{T_{\rm DR}}{m_\chi}\right)^{n+4}\right].
\ee
As long as the heating rate obeys $\Gamma_{\rm heat} \gg \mathcal{H}$, the solution to Eq.~\eqref{eq:DM_temperature_evol_final} is $T_\chi \simeq T_{\rm DR}$. In the opposite limit $\Gamma_{\rm heat} \ll \mathcal{H}$, the DM cools adiabatically with $T_\chi \propto a^{-2}$. For a heating rate of the form in Eq.~\eqref{eq:Gamma_heating} an analytic solution is possible \cite{2007JCAP...04..016B}. For the special case where the heating rate has the same redshift dependence as the Hubble expansion rate ($\Gamma_{\rm heat}/\mathcal{H} = \text{constant}$), Eq.~\eqref{eq:DM_temperature_evol_final} admits the solution,
\be
T_\chi = \frac{\Gamma_{\rm heat}/\mathcal{H}}{1+\Gamma_{\rm heat}/\mathcal{H}} T_{\rm DR}.
\ee
This regime is interesting since it allows $T_\chi \ll T_{\rm DR}$ while retaining the scaling $T_\chi \propto a^{-1}$. A concrete model realizing this regime was recently proposed in Ref.~\citep{Buen-Abad:2015ova}. The sound speed given in Eq.~\eqref{eq:DM_sound_speed} then takes the form
\be\label{eq:DM_sound_speed_final}
c_{\chi}^2 = \frac{T_\chi}{m_\chi}\left(\frac{5}{3}-\frac{\Gamma_{\rm heat}(T_{\rm DR})\left(T_{\rm DR} - T_\chi\right)}{3 \mathcal{H} T_\chi}\right).
\ee
We note that the above sound speed is generally very small for nonrelativistic DM ($T_\chi \ll m_\chi$) and thus has very little impact on the evolution of DM density fluctuations, except on very small length scales. We also note that in the limit $\Gamma_{\rm heat} \gg \mathcal{H}$, the term in the bracket in Eq.~\eqref{eq:DM_sound_speed_final} approaches $4/3$, leading to $c_\chi^2\rightarrow (4T_\chi/3m_\chi)$.
\subsubsection{Perturbation evolution}
We now turn our attention to computing the right-hand side of Eq.~\eqref{eq:DM_velocity}. It is important to notice that the momentum appearing in the integrand is the incoming DM momentum, while that appearing in the collision term given in Eq.~\eqref{final_coll_term} is the incoming DR momentum. We can use conservation of momentum to write \citep{Dodelson-Cosmology-2003}
\be
\int \frac{d^3p_2}{(2\pi)^3} \frac{\pp{2}}{m_\chi} C_{\chi(\pp{2})\tilde{\gamma}\leftrightarrow\chi\tilde{\gamma}}[\pp{2}] = -\frac{\eta_{\rm DR}}{\eta_\chi}\int \frac{d^3p_1}{(2\pi)^3} \hat{\bf p}_1 C_{\tilde{\gamma}(\pp{1})\chi\leftrightarrow\chi\tilde{\gamma}}[\pp{1}], 
\ee
where on the left-hand side, $\pp{2}$ is the incoming DM momentum, while on the right-hand side, $\pp{1}$ is the momentum of the incoming DR. With the help of this identity, we can then use Eq.~\eqref{final_coll_term} to compute the right-hand side of the DM velocity equation:
\be
\frac{-a}{16\pi m_\chi^3} \eta_{\rm DR}  \int \frac{d^3p_1}{(2\pi)^3} p_1^2(i\vec{k}\cdot\hat{\bf p}_1)\frac{\pa \fR^{(0)}(p_1)}{\pa p_1} \Big[\frac{1}{4}\sum_{l=1}^{\infty}(-i)^l(2l+1)\nu_l(p_1)P_l(\mu)\left(A_0(p_1)-A_l(p_1)\right)-P_1(\mu)\left(A_0(p_1) - A_1(p_1)\right)  v_\chi \Big].
\ee
Since $i\vec{k}\cdot\hat{\bf p}_1 = ik\mu = ikP_1(\mu)$, the angular integration is straightforward and yields
\be
-a\frac{1}{16\pi m_\chi^3} \frac{\eta_{\rm DR} }{3}  \Big[\theta_{\rm DR} -\theta_\chi \Big]\int \frac{p_1^2dp_1}{2\pi^2} p_1^2 \frac{\pa \fR^{(0)}(p_1)}{\pa p_1} \left(A_0(p_1)-A_1(p_1)\right).
\ee
We thus define the DM drag opacity
\be
 \dot{\kappa}_\chi  \equiv a\frac{1}{16\pi m_\chi^3} \frac{\eta_{\rm DR} }{3} \int \frac{p_1^2dp_1}{2\pi^2} p_1^2 \frac{\pa \fR^{(0)}(p_1)}{\pa p_1} \left(A_0(p_1)-A_1(p_1)\right) =\frac{4 \rho_{\rm DR}}{3 n_\chi^{(0)} m_\chi} \dot{\kappa}_{\rm DR-DM},
\ee
where we used the definition of the DR opacity given Eq.~\eqref{eq:dark_rad_opacity_def} in the last equality.

%
%\be
%a\frac{1}{8\pi m_\chi^3} \frac{\eta_{\rm DR}}{3}  \Big[\theta_{\rm DR} -\theta_\chi \Big]\int \frac{dp_1}{(2\pi)^2} \fR^{(0)}(p_1)   \left[4p_1^3 \left(A_0(p_1)-A_1(p_1)\right)+p_1^4\left(\frac{\pa A_0(p_1)}{\pa p_1}-\frac{\pa A_1(p_1)}{\pa p_1}\right)\right].
%\ee
%
%Using the definition of the DR opacity given in Eq.~\eqref{eq:dark_rad_opacity_def}, we can write the above as
%
%\be\label{eq:def_DM_drag_opacity}
%\left\{\frac{4 \rho_{\rm DR}}{3 n_\chi^{(0)} m_\chi} \dot{\kappa}_{\rm DR-DM} -a\frac{1}{48\pi m_\chi^3} \eta_{\rm DR} \int \frac{dp_1}{2\pi^2} \fR^{(0)}(p_1)   \left[p_1^4\left(\frac{\pa A_0(p_1)}{\pa p_1}-\frac{\pa A_1(p_1)}{\pa p_1}\right)\right]\right\} \Big[\theta_\chi -\theta_{\rm DR} \Big]\equiv \dot{\kappa}_\chi  \Big[\theta_\chi -\theta_{\rm DR} \Big],
%\ee
%
In summary, the DM equations take the form
\ba
\dot{\de}_\chi+\theta_\chi-3\dot{\phi}&=&0\label{eq:DM_continuity_right_not},\\
\dot{\theta}_\chi -c_{\chi}^2k^2\de_\chi+\mathcal{H} \theta_\chi - k^2\psi &=&\dot{\kappa}_\chi \left[\theta_\chi -\theta_{\rm DR} \right]\label{eq:DM_euler_approx_sec}.
\ea
We observe that the details of the DM particle model only enter through the functions $\dot{\kappa}_\chi$ and $c_\chi^2$. It is thus clear that two models predicting the same values for these functions will lead to a very similar structure formation scenarios. This is the basic idea behind the ETHOS framework.

\section{Impact of elastic dark matter self-interaction on the evolution of linear cosmological perturbations}\label{sec_app:chichi_int}
%%%%
In this Appendix, we briefly consider the physical reasons why elastic DM self-interaction $\chi\chi\leftrightarrow\chi\chi$ is irrelevant to the cosmological evolution of \emph{linear} perturbations for nonrelativistic DM. As we discuss below, this is essentially a consequence of energy and momentum conservation.  We emphasize that the following discussion is only appropriate for the case of \emph{elastic} DM collisions. Inelastic DM collisions (see e.g. Refs.~\cite{TuckerSmith:2001hy,Finkbeiner:2007kk,ArkaniHamed:2008qn,Graham:2010ca,2011PhRvL.106q1302L,McCullough:2013jma,Schutz:2014nka}) can in general affect the evolution of linear DM perturbations, but it is beyond the scope of the current paper to explore this possibility. Restricting ourselves to the elastic case, we first consider the impact of the $C_{\chi\chi\leftrightarrow\chi\chi}$ collision integral on the evolution of DM number density fluctuations. Intuitively, the process $\chi\chi\leftrightarrow\chi\chi$ preserves the number of $\chi$ particles and thus cannot affect the evolution of the DM number density. This can also be shown mathematically by looking at the structure of the right-hand side of Eq.~\eqref{eq:DM_monopole_first_eq} once DM self-interaction is included. The important term has the generic form (neglecting prefactors irrelevant to our analysis)
\ba
\int \frac{d^3p_1}{(2\pi)^3}\frac{1}{E_1}C_{\chi\chi\leftrightarrow\chi\chi}\left[\pp{1}\right] =\hspace{12cm} \\
 \int d\Pi_1 d\Pi_2 d\Pi_3 d\Pi_4 \left(\frac{1}{\eta_\chi}\sum_{\rm states}|\mathcal{M}|^2\right)(2\pi)^4\de^4(P_1+P_2-P_3-P_4)\left(\fD(\pp{3})\fD(\pp{4})-\fD(\pp{1})\fD(\pp{2})\right),\nonumber
\ea
where $d\Pi_i$ is the Lorentz invariant phase-space integral measure [see Eq.~\eqref{eq:integral_measure}], $P_i$ stands for the $i^{\rm th}$ four-momentum, $\pp{i}$ is the $i^{\rm th}$ three-momentum, $p_i = |\pp{i}|$, $E_i^2 = m_\chi^2+p_i^2$, $m_\chi$ is the DM mass, $\eta_\chi$ is the DM degeneracy factor, $|\mathcal{M}|^2$ is the squared matrix element for the $\chi\chi\leftrightarrow\chi\chi$ process, and where $\fD$ is the DM phase-space distribution function. On the one hand, we  note that the integral measure, the squared matrix element, and the delta function are symmetric under the exchange of the initial and final states ($\pp{1}\leftrightarrow \pp{3}$ and $\pp{2}\leftrightarrow \pp{4}$). This is easily seen by writing the squared matrix element in terms of Mandelstam variables and by noting that the latter are invariant under the exchange of initial and final states for the $\chi\chi\leftrightarrow\chi\chi$ process. On the other hand, the term involving the phase-space distributions is antisymmetric under the exchange of the initial and final states. This immediately implies that the above collision integral has to exactly vanish and that DM self-interaction does not impact the evolution of DM density fluctuations. 

We now turn our attention to the evolution of DM velocity perturbations. In the presence of DM self-interaction, the right-hand side of Eq.~\eqref{eq:DM_velocity} will admit a term of the generic form,
\ba\label{eq:app_chichicoll_theta}
\eta_\chi  \int \frac{d^3p_1}{(2\pi)^3}\frac{\vec{k}\cdot\pp{1}}{E_1^2}C_{\chi\chi\leftrightarrow\chi\chi}\left[\pp{1}\right]= \hspace{12cm}\\
 \int d\Pi_1 d\Pi_2 d\Pi_3 d\Pi_4 \left(\sum_{\rm states}|\mathcal{M}|^2\right)(2\pi)^4\de^4(P_1+P_2-P_3-P_4)\left(\frac{\vec{k}\cdot\pp{1}}{E_1}\right)\left(\fD(\pp{3})\fD(\pp{4})-\fD(\pp{1})\fD(\pp{2})\right),\nonumber
\ea
where $\vec{k}$ is the wave number characterizing the cosmological perturbation. Since we are integrating over the phase space of all incoming and outgoing particles, the above expression is symmetric under the exchange $\pp{1}\leftrightarrow \pp{2}$ and $\pp{3}\leftrightarrow \pp{4}$. This implies the following equality:
\be\label{eq:app_eq1}
 \int d\Pi_1 d\Pi_2 d\Pi_3 d\Pi_4 \left(\sum_{\rm states}|\mathcal{M}|^2\right)(2\pi)^4\de^4(P_1+P_2-P_3-P_4)\left(\vec{k}\cdot(\frac{\pp{1}}{E_1}-\frac{\pp{2}}{E_2})\right)\left(\fD(\pp{3})\fD(\pp{4})-\fD(\pp{1})\fD(\pp{2})\right)=0.
 \ee
 We also note that momentum conservation implies the following equality: 
\ba\label{eq:ap_show_chichi_theta}
 \int d\Pi_1 d\Pi_2 d\Pi_3 d\Pi_4 \left(\sum_{\rm states}|\mathcal{M}|^2\right)(2\pi)^4\de^4(P_1+P_2-P_3-P_4)\left(\frac{\vec{k}\cdot(\pp{1}+\pp{2})}{m_\chi}\right)\left(\fD(\pp{3})\fD(\pp{4})-\fD(\pp{1})\fD(\pp{2})\right)\qquad\\
\,\,= \int d\Pi_1 d\Pi_2 d\Pi_3 d\Pi_4 \left(\sum_{\rm states}|\mathcal{M}|^2\right)(2\pi)^4\de^4(P_1+P_2-P_3-P_4)\left(\frac{\vec{k}\cdot(\pp{3}+\pp{4})}{m_\chi}\right)\left(\fD(\pp{3})\fD(\pp{4})-\fD(\pp{1})\fD(\pp{2})\right).\nonumber
\ea
However, under the exchange of the initial and final states ($\pp{1}\leftrightarrow \pp{3}$ and $\pp{2}\leftrightarrow \pp{4}$), the first line of Eq.~\eqref{eq:ap_show_chichi_theta} becomes equal to minus the second line, which immediately implies that 
\be\label{eq:app_eq2}
 \int d\Pi_1 d\Pi_2 d\Pi_3 d\Pi_4 \left(\sum_{\rm states}|\mathcal{M}|^2\right)(2\pi)^4\de^4(P_1+P_2-P_3-P_4)\left(\frac{\vec{k}\cdot(\pp{1}+\pp{2})}{m_\chi}\right)\left(\fD(\pp{3})\fD(\pp{4})-\fD(\pp{1})\fD(\pp{2})\right)=0
 \ee
Reconciling Eqs.~\eqref{eq:app_eq1} and \eqref{eq:app_eq2} in the nonrelativistic limit ($E\simeq m_\chi$) immediately requires that the collision integral given in Eq.~\eqref{eq:app_chichicoll_theta} exactly vanishes. The elastic $\chi\chi\leftrightarrow\chi\chi$ process thus cannot affect the evolution of the DM velocity perturbations. We finally note that if one were to consider higher moments of the DM Boltzmann equation, there are no \emph{a priori} conservation laws that prohibit the existence of nonvanishing collision integrals. However, these higher DM moments are highly suppressed for nonrelativistic DM, and thus play essentially no role in the evolution of linear DM perturbations.

%%%
\bibliography{dark_matter_ref}
%%%%

%%%%%
\end{document}